\RequirePackage{fix-cm}

\documentclass[smallextended]{svjour3}       % onecolumn (second format)
\smartqed  % flush right qed marks, e.g. at end of proof
\usepackage{graphicx}
\usepackage{amsmath}
\usepackage{bm}
\usepackage{cases}
\begin{document}

\title{Modeling Coating Flow and Surfactant Dynamics inside the Alveolar Compartment%\thanks{Grants or other notes
}

\titlerunning{Surfactant Dynamics in Alveoli}        % if too long for running head

\author{D. Kang         \and
        M. Chugunova \and
        A. Nadim  \and
        A. J. Waring \and        
        F. J. Walther
        %etc.
}


\institute{D. Kang  \at
             Institute of Mathematical Sciences,
             Claremont Graduate University, Claremont, CA 91711\\            
              \email{di.kang@cgu.edu}           %  \\
           \and
          M. Chugunova \at
             Institute of Mathematical Sciences,
             Claremont Graduate University, Claremont, CA 91711\\            
                  \email{marina.chugunova@cgu.edu}  
          \and
          A. Nadim \at
              Institute of Mathematical Sciences,
              Claremont Graduate University, Claremont, CA 91711\\            
                    \email{ali.nadim@cgu.edu}
          \and
           A. J. Waring \at
                                Los Angeles Biomedical Research Institute, Harbor-University of California at Los Angeles (UCLA) Medical Center, Torrance, CA 90502 \\
                                \email{awaring@labiomed.org}
           \and
           F. J. Walther \at
           Los Angeles Biomedical Research Institute, Harbor-University of California at Los Angeles (UCLA) Medical Center, Torrance, CA 90502 \\
           \email{fwalther@labiomed.org}                
}

\date{Received: date / Accepted: date}

\maketitle

\begin{abstract}
 We derive a new model for the coating flow inside the alveolar compartment, taking into account pulmonary surfactant production and recycling by Type 2 cells as well as its degradation.  As the thickness of alveolar coating is much smaller than the average radius of the alveoli, we employ the classical lubrication approximation to describe the thin liquid film dynamics in the presence of pulmonary surfactant, which is a surface tension reducing agent and thus prevents the lungs from collapse. In the lubrication limit, we derive a degenerate system of two coupled parabolic partial differential equations that describe the time evolution of the thickness of the coating film inside the alveoli together with that of the surfactant concentration at the interface.  We present numerical simulations using parameter values consistent with experimental measurements.
\end{abstract}

\section{Introduction}

\begin{figure}
\label{fig1}
\begin{center}
\includegraphics[height=3.5cm] {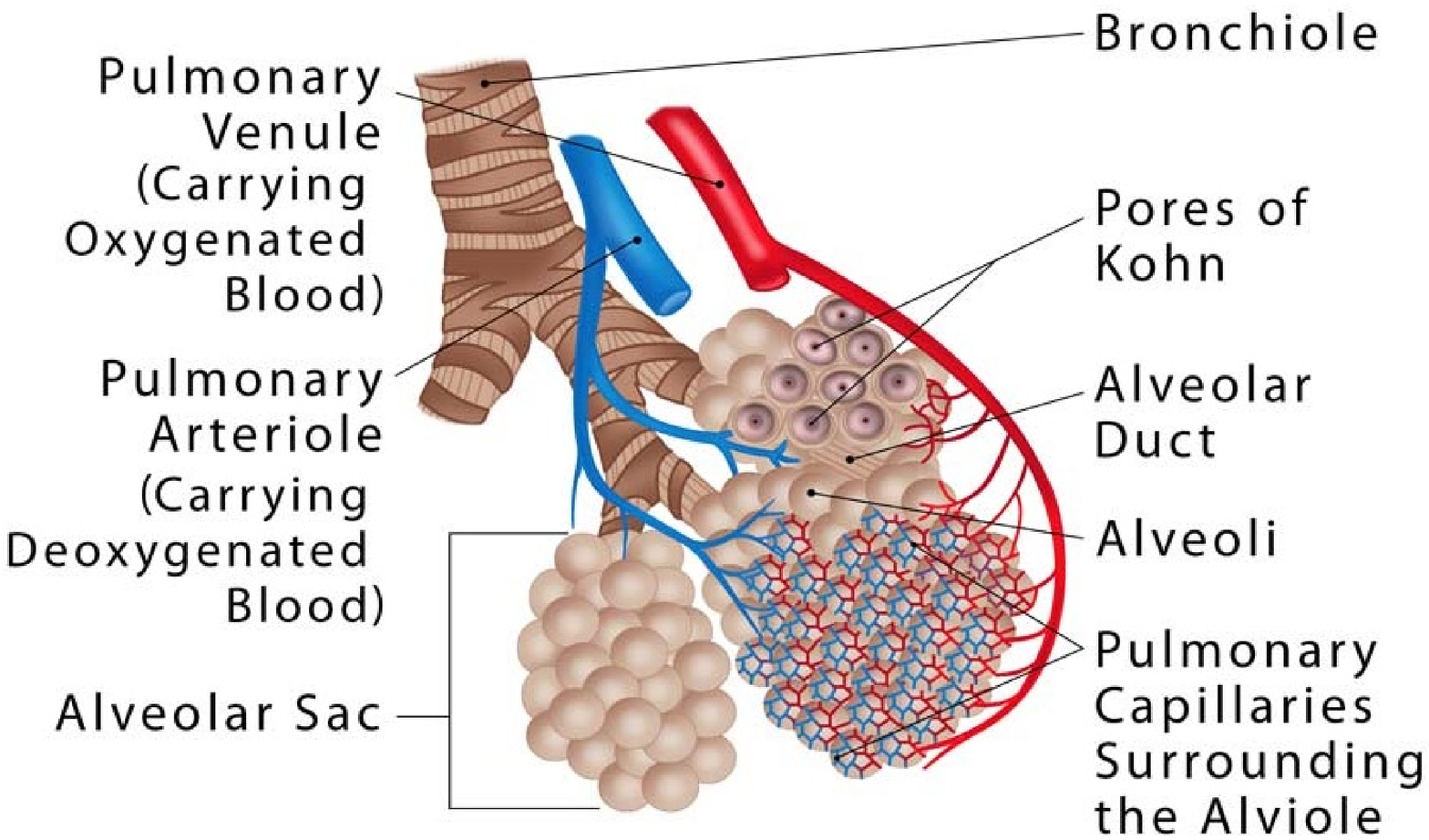} \hspace{1cm}
\includegraphics[height=3.5cm] {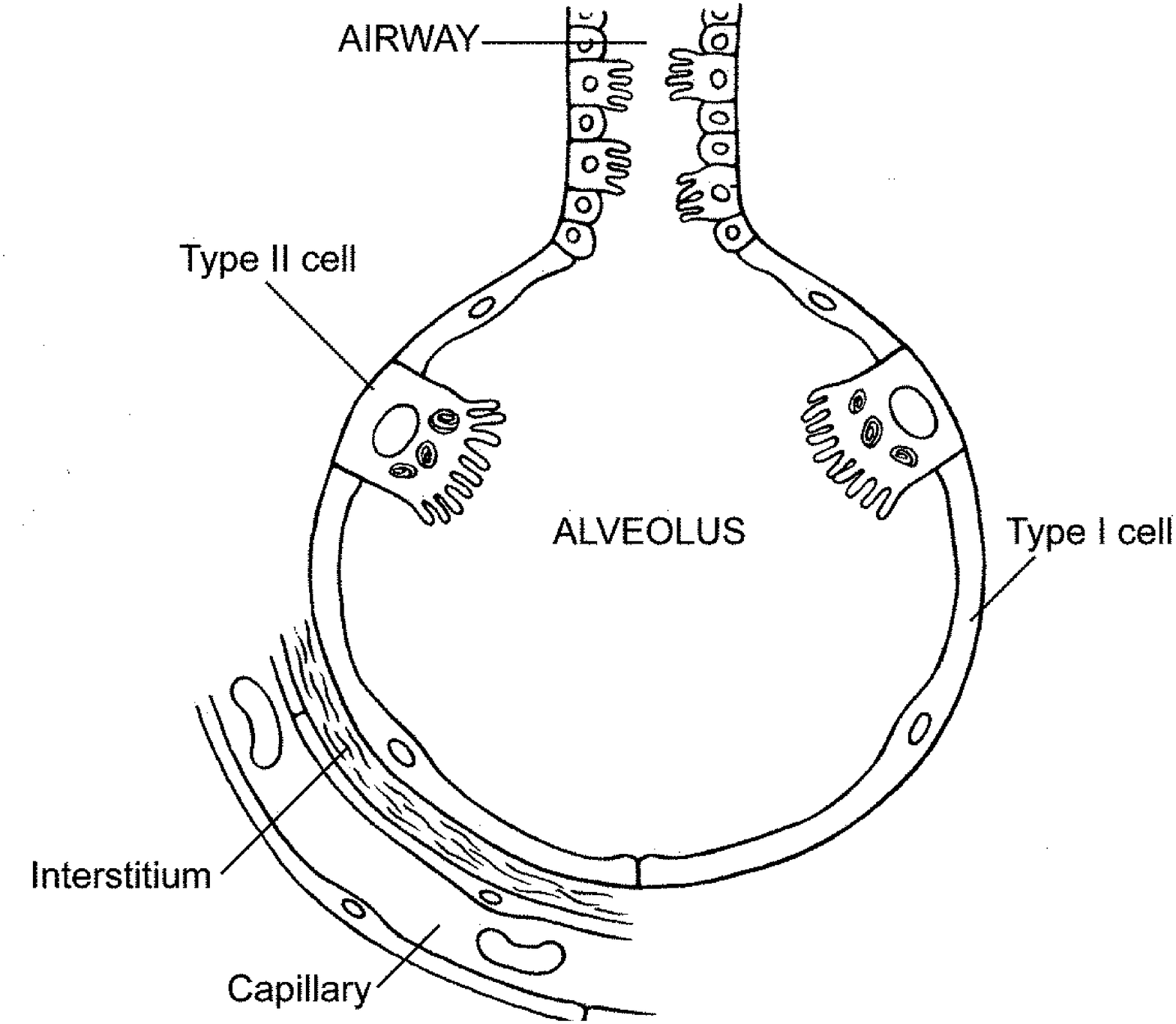}
\end{center}
\caption{Alveolar sac schematic on the left [adapted from www.therespiratorysystem.com] and cell types present in alveolar compartment on the right [adapted from Michael et al.~(2002)].}
\end{figure}

Oxygen exchange in the lungs occurs across the membranes of small balloon-like structures called alveoli attached to the branches of the bronchial passages. These alveoli inflate and deflate with inhalation and exhalation. The behavior of the alveoli is largely dictated by Laplace's law (that describes the pressure difference across an interface in fluid mechanics) involving surface tension. It takes some effort to breathe in because these tiny balloons must be inflated, but the elastic recoil of the tiny balloons assists us in this process. Inflating the alveoli in the process of respiration requires an excess pressure inside the alveoli relative to their surroundings. This is accomplished by making the pressure in the thoracic cavity negative with respect to atmospheric pressure. The amount of net pressure required for inflation is dictated by the surface tension and radii of the tiny balloon-like alveoli. During inhalation the radii of the alveoli increase from about 0.05~mm to 0.1~mm. The alveolar lining fluid (the alveolar hypophase, without surfactant)  has a nominal surface tension of about 50~dyn/cm so the required net outward pressure is about $\Delta P = 15$~mmHg when $r=0.05$~mm and $\Delta P = 7.5$~mmHg when $r= 0.1$~mm, but the actual normal pressure difference in respiration is only about 1~mmHg \cite{Shier2007}.

A remarkable property of lung surfactant, which coats the alveoli, is that it reduces the surface tension by a factor of about 15 so that a 1~mmHg pressure differential is sufficient to inflate the alveoli. There appears to be a nearly constant amount of this surfactant per alveolus, so that when the alveoli are deflated it is more concentrated on the surface. Since the surface-tension-lowering effect of surfactant depends on this concentration, it diminishes the required pressure for inflation of the alveoli at their most critical phase. For a given surface tension, the pressure to inflate a smaller bubble is greater. It is the surfactant which makes it possible to inflate the alveoli with only about 1~mmHg of excess pressure over the surroundings. Because neighboring alveoli communicate with each other via alveolar pores (pores of Kohn connect alveoli to adjacent alveoli), their surface tensions must be different (if they are different in size) in order to prevent the collapse of small alveoli in favor of large ones. Mechanical coupling of alveoli via the interstitial tissue is an additional mechanism that prevents alveolar collapse \cite{Mason2006}.

The lung is very sensitive to gravity, and it is of current interest to know how its function is altered in the weightlessness of space. For example, studies on NASA Spacelabs \cite{space1} show an unexpectedly large increase in the diffusing capacity of the alveolar membrane due to uniform alveolar expansion. Effect of microgravity on pulmonary surfactant properties was studied in \cite{space2}. The properties of pulmonary surfactant were assessed by the evaluation of surface activity (surface tension) and the experiments indicated that there was roughly a 50\% decrease in surface activity of pulmonary surfactant films after some time in the microgravity environment.

%\marginpar{Check the units of the volumes and sizes; microns?}
The epithelium lining the walls of the alveoli is composed primarily of large, squamous Type 1 cells and smaller, granular and roughly cuboidal Type 2 cells (see Figure \ref{fig1}).  The alveolar Type 1 cell (volume 1,800~$\mu$m$^3$) is an important component of the air-blood barrier, as it overlies capillaries in the alveolar wall and comprises most (90\%) of its surface area. This has been described in papers about proteins that are predominantly expressed by Type 1 cells in the lungs, such as the plasma membrane protein T1-a \cite{Williams} and aquaporins, a family of water channels \cite{Verkman}. Although Type 2 cells (volume 900~$\mu$m$^3$ and thickness 0.2~$\mu$m) occupy only 10\% of the alveolar surface area, there are about twice as many of them (60\% of alveolar epithelial cells are Type 2) as Type 1 cells \cite{Crapo,Haies}. At the same time Type 2 cells comprise only 15\% of peripheral lung cells; they are typically found at the alveolar-capillary barrier, and they have an apical surface area of about 250~$\mu$m$^2$ per cell \cite{Mason2006}. Alveolar surfactant has a half life of about 24 hours, once secreted, depending on maturity/illness according to studies with stable isotopes \cite{Carnielli}. Viscosity values for some clinical animal-derived lung surfactants are presented in Table 1 of \cite{Frans}. 
One of the major functions of the Type 2 cells is synthesis and secretion of surfactant. Cryoscanning electron microscopy of frozen tissue demonstrated surfactant to cover extended areas of alveolar surface as a continuous, thin layer \cite{Heinz}. Approximately 90\% of the surfactant is recycled from the alveolar space back into the same Type 2 cells \cite{castranova1988}. Type 2 cells are called defenders of the alveolar epithelium because they proliferate to restore Type 1 cells after lung damage. Alveolar Type 1 and 2 cells transport sodium to keep the alveolus relatively free from fluid, and participate in innate immunity to fight infection.

The baby's first breath depends a lot upon the presence of pulmonary surfactant and is made more difficult in premature infants by the incomplete formation of this surface-tension-reducing agent.  Approximately 7\% of all infants are born prematurely and half of them develop respiratory distress syndrome due to lack of the surfactant. Aerosol delivery of synthetic lung surfactant is being investigated as a new technique of respiratory support for such babies \cite{walther2014} and as a potential replacement for the widely-used intra-tracheal delivery of animal-derived surfactant. One of the objectives of our research is to better understand the required properties of substitute synthetic surfactant through mathematical modeling of lung surfactant dynamics.

Biomedical applications of surfactant dynamic models also include  surfactant-based drug delivery systems. Consider the delivery pathway of a liquid medication drop as it makes its way from the trachea to the alveoli. It starts as a liquid plug, progresses to a deposited film lining the airways, establishes a surface layer, and finally reaches the alveolar compartment. The surface tension of a surfactant-coated layer depends on the local concentration of surfactant, which changes owing to flow, surface deformation and surface diffusion of the molecule. Surface tension gradients caused by variations of surfactant concentration give rise to Marangoni flows that distribute the surfactant \cite{Halp1997,Halp2008}.

Flows of thin liquid films influenced by different types of surfactant constitute a complex area of research with various contributions by chemists, physicists, engineers and mathematicians. The analysis of surface tension and surface active agents in the dynamics of thin viscous liquid films is of interest in many applications in industrial coating, biomedical fields or film drainage in emulsions and foams. Surfactants acting at the interface of a fluid film induce surface tension gradients that influence the dynamics of the fluid film. At the same time, the surfactant itself spreads along the interface due to the flow arising from those surface tension gradients. The latter aspect is called the Marangoni effect.

The lubrication approximation is the classical approach for studying the dynamics of thin viscous films. In spherical geometry, which is the one we adopt in our work, well-posedness of the thin film model was analyzed in \cite{sphere1,sphere2}. The motion of a Newtonian viscous fluid layer on a solid horizontal plane, with a monolayer of insoluble surfactant on its surface was modeled by Jensen and Grotberg \cite{JG} resulting in:
\begin{numcases}{}
h_t  + \tfrac{1}{3}( h^3( \mathcal{S} h_{xxx} - \mathcal{G} h_x +
 3 \mathcal{A} h^{-4} h_x))_x + \tfrac{1}{2}(h^2 \sigma_x)_x = 0, \qquad \hfill \label{I:1}\\
\Gamma_t  + \tfrac{1}{2}( \Gamma h^2( \mathcal{S} h_{xxx} -
\mathcal{G} h_x + 3 \mathcal{A} h^{-4} h_x))_x +(\Gamma h
\sigma_x)_x = (\mathcal{D}(\Gamma)
\Gamma_{x})_x, \qquad \hfill \label{I:2}
\end{numcases}
where $h$ is the film height, $\Gamma$ is the surfactant
concentration in the monolayer, $\sigma(\Gamma)$ is the surface
tension which depends on surfactant concentration, $\mathcal{S}$ is a parameter related to the capillary forces (i.e., surface tension), $\mathcal{G}$ is the parameter characterizing the gravitational force directed vertically downwards, $\mathcal{A}$ is related to the
Hamaker constant and connected with intermolecular van der Waals
forces, and $\mathcal{D}$ is related to the surface diffusivity of surfactants. 
This model is the generalization of the
original system derived by Gaver and Grotberg in 1990 \cite{GG} and studied in \cite{surf1,surf2};
the new model includes a nonlinear equation of state and van der
Waals forces. Capillarity is an important factor in stabilizing
the film against instabilities due to van der Waals forces. The
latter can cause a film to
rupture. Since film rupture has such a dramatic effect on the
spreading process, it interferes severely with methods of delivery
of surfactant or drugs, making it of considerable importance to
establish the conditions under which rupture might occur.
Marangoni forces alone are not sufficient to induce the film
thickness to reach zero in finite time, but they can deform the
film to allow van der Waals forces to overcome the stabilizing
effects of capillarity (and of the surfactant monolayer itself)
and give rise to rupture and dry-out.

Various models are in use relating surface tension $\sigma$ to surfactant concentration $\Gamma$.
Under constant temperature, a fundamental equation of chemical
thermodynamics relates the concentration dependent surface tension
$\sigma$ to the free energy, $\Phi$, and the chemical potential,
$\Phi'$, where both functions depend on the surfactant concentration
$\Gamma$:
\begin{equation}\label{I:4}
\sigma(\Gamma) = \Phi(\Gamma) - \Gamma \Phi'(\Gamma).
\end{equation}
By convexity of the free energy this relation implies a monotone
decrease of surface tension for nonnegative concentration.
It was confirmed by numerous experiments that the surface diffusivity of surfactant
is not a constant \cite{EHLW,EHLW10,GW}, and can be modeled by a nonlinear function of the
surfactant concentration $\Gamma$ (see e.g., \cite[(6.1) and
(6.2), pp.~158-159]{BG}), namely,
\begin{equation}\label{I:5}
\sigma(\Gamma) = (1 + \theta \Gamma)^{-3}\,, \qquad \mathcal{D}(\Gamma) = (1
+ \tau \Gamma)^{-k},
\end{equation}
where $\theta,\, \tau$ and $k$ are some positive empirical parameters.
In reality, the parameter $\theta$ depends on the material properties
of the monolayer  (cf.~\cite{BG} for details). The empirical
relation (\ref{I:5}) is based on experimental data obtained for
the inner subinterval  $0 < \delta \leq \sigma \leq 1$.
For example, if $\theta = 0.15$ then (\ref{I:5})
well describes an oil layer on water \cite[Figure 2, p.~159]{BG} .
It is very difficult to obtain high experimental accuracy when the surfactant concentration reaches near saturation levels.

In many applications the dependence $\sigma(\Gamma)$
is taken to follow the Frumkin equation of state (cf.~\cite[(16),
p.~324]{Luc} for example):
\begin{equation}\label{I:Frum} \sigma(\Gamma) = \sigma_0 + 2.303 R\,T\,
\Gamma_s \bigl( b ({\Gamma}/{\Gamma_s})^2 +  \ln (1-
{\Gamma}/{\Gamma_s}) \bigr),
\end{equation}
where $\sigma_0$ is the surface tension of pure solvent and $b$ is
the Frumkin constant. 
%(for example, $b = 0$ for pentanoic acid) % Not sure if we need this
This equation, first formulated as an empirical relation, can be obtained
from a general surface equation of state %(Lucassen-Reynders, 1967)
if one assumes ideal surface behavior (i.e., surface activity
coefficients close to unity).

In this paper we model and study the behavior of surfactant driven thin film flow inside the alveoli whose shapes are considered to be approximately spherical, taking into account pulmonary surfactant production and recycling by Type 2 cells as well as its
degradation. We model production of the surfactant by introducing a position dependent source term and we assume a uniform degradation rate for the surfactant. We also allow for the inflation and deflation of the alveolar compartment by taking the radius of the alveolus to be time periodic.

The article is structured as follows: in Section 2 we apply the lubrication approximation to derive and simplify the mathematical model; in Section 3 we carry out numerical simulations for the constant radius
case; and in Section 4 we present the numerical simulation results for the time periodic case. Section 5 summarizes the results and discusses their significance.

% SECTION

\section{Model Formulation}

In this section we derive a mathematical model for the dynamics of the alveolar lining fluid on the inner surface of a spherical alveolus and the insoluble surfactant on its liquid-air interface. This derivation is analogous to our earlier work \cite{kang2016,kang2017} which considered thin films on the outside of a sphere undergoing rotation or with thermal gradients. Consider a thin viscous liquid film on the inner surface of a sphere of time dependent radius $R(t)$ in the presence of gravity and surface tension. We assume that the system is axisymmetric with respect to the gravitational axis. Let $r$ denote the distance from origin at the center of the sphere, and $\theta$ the polar angle relative to the positive vertical axis. There is insoluble surfactant on the liquid-air interface, which is produced at rate $\alpha(\theta)$ which is related to the distribution of Type 2 alveoli cells and degrades with a constant rate constant $\beta$. The thickness of the thin liquid film is denoted by $h(\theta,t)$ and the concentration of surfactant is denoted by $\Gamma(\theta,t)$. The interface of the thin film $r=R(t)-h(\theta,t)$ is the zero-level set of function
\[
{\mathcal F}(r,\theta,t)=r-R(t)+h(\theta,t)\,.
\]
The kinematic boundary condition $D{\mathcal F}/Dt=0$ reads
\begin{equation}
\frac{\partial h}{\partial t}=-v_{r}-\frac{v_{\theta}}{r}\frac{\partial h}{\partial\theta}+\dot{R}(t)\label{eq:kinematic BC}
\end{equation}
at the liquid-air interface $r=R(t)-h(\theta,t)$.

The continuity equation in spherical coordinates can be written as
\[
\frac{1}{r^{2}}\frac{\partial}{\partial r}\left(r^{2}v_{r}\right)+\frac{1}{r\sin\theta}\frac{\partial}{\partial\theta}\left(v_{\theta}\sin\theta\right)=0\,.
\]
Multiply this equation by $r^{2}\sin\theta$ and integrate along $r$
at a fixed $\theta$ to get
\begin{equation}
\sin\theta \left. (r^{2}v_{r})\right|_{R-h}^{R}+\frac{\partial}{\partial\theta}\int_{R-h}^{R}rv_{\theta}\sin\theta dr-\left. \left(rv_{\theta}\sin\theta\right)\right|_{R-h}\frac{\partial h}{\partial\theta}=0.
\end{equation}
Combining this with the kinematic boundary condition, along with the no-slip boundary condition $v_{r}=\dot{R}(t)$ at $r=R(t)$, we derive the time evolution equation
\begin{equation}
(R-h)^{2}\frac{\partial h}{\partial t}+\frac{\partial}{\partial\theta}\left(\int_{R-h}^{R}rv_{\theta}dr\right)+\dot{R}(2R h-h^2)=0.\label{eq:evolution eqn}
\end{equation}
In order to get a self-contained partial differential equation for $h(\theta,t)$, we need to relate the velocity component $v_{\theta}$ to the film thickness.

We consider the standard lubrication form of the momentum equations
from the Navier-Stokes equations (namely that the hydrostatically modified pressure is uniform across the film and the gradient of that pressure in the long direction balances the dominant viscous term), having the forms
\begin{equation}
\frac{\partial P}{\partial r}=0,\label{eq:r-mom}
\end{equation}
\begin{equation}
\frac{1}{r}\frac{\partial P}{\partial\theta}=\frac{\mu}{r^{2}}\frac{\partial}{\partial r}\left(r^{2}\frac{\partial v_{\theta}}{\partial r}\right),\label{eq:theta-mom}
\end{equation}
where $P$ denotes the modified pressure field, which is defined by
\[
P=p-\rho\,\mathbf{g\cdot x}= p+\rho g \, r \cos\theta \,.
\]
The normal stress balance at the liquid-air interface $r=R-h$ states
that the pressure in the film needs to equal to the air pressure $p_{0}$
plus the capillary contribution given by $\sigma\nabla\cdot\mathbf{n}$,
where $\sigma$ is the surface tension dependent upon
the surfactant concentration $\Gamma(\theta,t)$ and $\mathbf{n}$
is the normal unit vector pointing towards the air phase. Under the lubrication approximation, the normal stress balance simplifies to
\begin{align}
P|_{r=R-h}&=p_{0}+\sigma\nabla\cdot\mathbf{n}\nonumber \\
&=p_{0}-\sigma\left(\frac{2}{R}+\frac{1}{R^{2}}\left(2h+\frac{1}{\sin\theta}\frac{\partial}{\partial\theta}\left(\sin\theta\frac{\partial h}{\partial\theta}\right)\right)\right).\label{eq:modified pressure}
\end{align}
From the $r$-momentum equation (\ref{eq:r-mom}) we know that the modified
pressure $P$ is independent of $r$, thus we have
\[
P(\theta,t)=p_{0}+\sigma\nabla\cdot\mathbf{n}+\rho g(R-h)\cos\theta.
\]
By integrating the $\theta$-momentum equation twice, the general solution for $v_{\theta}$ is found to be
\begin{equation}
v_{\theta}=\frac{1}{2\mu}\frac{\partial P}{\partial\theta}r-\frac{C_{1}}{r}+C_{2}\,.
\end{equation}
The integration constants $C_{1}$ and $C_{2}$ can be obtained from the no-slip boundary condition at $r=R$ and the tangential stress balance at $r=R-h$:
\[
\mathbf{n}\cdot(\bm{\tau}_{2}-\bm{\tau}_{1})\cdot\mathbf{t}=\nabla_{s}\sigma\,.
\]
Here $\bm{\tau}_1$ and $\bm{\tau}_2$ are the viscous stresses in the liquid and air phases, respectively, the latter being negligible compared to the former. In our model, due to the presence of surfactant, surface tension $\sigma$ is a function of surfactant concentration $\Gamma(\theta,t)$; thus $\sigma$ is also a function of $\theta$ and $t$. Expanding the tangential stress balance along the $\theta$
direction and simplifying under the lubrication approximation yield
\begin{equation}
\mu r\frac{\partial}{\partial r}\left(\frac{v_{\theta}}{r}\right)=\frac{1}{r}\frac{\partial\sigma}{\partial\theta},
\end{equation}
at the interface $r=R-h$. Applying this boundary condition and the no-slip condition at $r=R$ enable the integration constants to be obtained as
\begin{align*}
C_{1}&=-\frac{R^{2}(R-h)}{2\mu(R+h)}\frac{\partial P}{\partial\theta}+\frac{R(R-h)}{\mu(R+h)}\frac{\partial\sigma}{\partial\theta}, \\
C_{2}&=-\frac{R^{2}}{\mu(R+h)}\frac{\partial P}{\partial\theta}+\frac{(R-h)}{\mu(R+h)}\frac{\partial\sigma}{\partial\theta}\,.
\end{align*}

Upon inserting these into the expression for $v_\theta$ and evaluating its integral as it appears in the evolution equation (\ref{eq:evolution eqn}) and making use of the assumption that $h\ll R$, we obtain the leading-order governing equation for $h(\theta,t)$:
\begin{equation}
\frac{\partial h}{\partial t}-\frac{1}{R^{2}\sin\theta}\frac{\partial}{\partial\theta}\left(\frac{h^{3}\sin\theta}{3\mu}\frac{\partial P}{\partial\theta}-\frac{h^{2}\sin\theta}{2\mu}\frac{\partial\sigma}{\partial\theta}\right)+\frac{2h}{R}\dot{R}(t)=0
\label{eq:govern eqn for h}
\end{equation}
where
\begin{equation}
P(\theta,t)=\rho gR\cos\theta-\frac{\sigma}{R^{2}}\left(2h+\frac{1}{\sin\theta}\frac{\partial}{\partial\theta}\left(\sin\theta\frac{\partial h}{\partial\theta}\right)\right).
\end{equation}

The total mass or volume of the alveolar fluid is conserved, i.e.,
\begin{equation}
\frac{d}{dt}\int^{R}_{R-h}\int^{\pi}_{0}h(\theta,t) r^2 \sin\theta drd\theta=0 \,,
\end{equation}
which can be simplified after applying the lubrication approximation to the form:
\begin{equation}\label{eq:conservation for h}
\frac{d}{dt}\left(R^2(t)\int^{\pi}_{0}h(\theta,t)\sin\theta d\theta \right)=0 \,.
\end{equation}

Surface tension $\sigma$ depends on the surfactant concentration, $\sigma=\sigma(\Gamma)$, and $\Gamma$ satisfies the interface transport equation (e.g., see \cite{nadim})
%\marginpar{This needs a reference. I'll add a citation to my paper.}
\begin{equation}
\frac{D\Gamma}{Dt}+\left(\nabla_{s}\cdot\mathbf{ v_{s}}\right)\Gamma+\left(\text{\ensuremath{\nabla}}_{s}\cdot\mathbf{n}\right)v_{n}\Gamma=\nabla_{s}\cdot\left(D_{s}\nabla_{s}\Gamma\right)+\alpha-\beta \Gamma, \label{eq:full eqn for Gamma}
\end{equation}
where $\alpha$ decribes the production of surfactant, $\beta$ is the degradation rate, $D/Dt=\partial/\partial t+\mathbf{v_{s}}\cdot \nabla_{s}$, $v_{n}=\mathbf{v}\cdot\mathbf{n}$ and $\nabla_{s}$ is the surface gradient which can be defined as $(\mathbf{I}-\mathbf{nn})\cdot\nabla$.

%\marginpar{Since normal n points into the air, shouldn't n=-er and vn=-Rdot?}
Under the lubrication assumptions, we can approximate $\mathbf{v_{s}}\approx v_{\theta}\mathbf{e_{\theta}}$, $v_{n}\approx -\dot{R}$ and $\mathbf{n}\approx -\mathbf{e_{r}}$. Thus we can simplify Equation (\ref{eq:full eqn for Gamma}) as
\begin{equation}
\frac{\partial \Gamma}{\partial t}+\frac{1}{R\sin\theta}\frac{\partial}{\partial \theta}(v_{\theta} \Gamma\sin\theta)+\frac{2\dot{R}}{R}\Gamma=\frac{1}{R^2 \sin\theta}\frac{\partial}{\partial \theta}\left(\sin\theta D_{s}\frac{\partial \Gamma}{\partial \theta}\right)+\alpha(\theta)-\beta \Gamma\,.
\end{equation}
Substituting the expression for $v_{\theta}$  at $r=R-h$, and using the lubrication approximation $h\ll R$, we obtain
\begin{multline}
\frac{\partial\Gamma}{\partial t}+\frac{1}{\sin\theta}\frac{\partial}{\partial\theta}\left(\frac{\sin\theta}{\mu R^2}\left(\frac{\partial\sigma}{\partial\theta}h-\frac{1}{2}\frac{\partial P}{\partial \theta} h^{2}\right)\Gamma\right)
\\=\frac{1}{\sin\theta}\frac{\partial}{\partial\theta}\Big(\frac{D_{s}}{R^{2}}\sin\theta\frac{\partial\Gamma}{\partial\theta}\Big)+ \alpha(\theta)-\Big(\beta+\frac{2\dot{R}}{R}\Big) \Gamma\,.
\label{eq:governing eqn for gamma}
\end{multline}
In the absence of production and degradation of the surfactant, i.e., when $\alpha=\beta=0$, the total amount of surfactant satisfies the total mass conservation:
\begin{equation}
\frac{d}{dt}\left( R^2(t) \int^{\pi}_{0}\Gamma(\theta,t)\sin\theta d\theta \right)=0\,.
\end{equation}
When the production and degradation of surfactant are present, the total amount of surfactant at steady state, $\Gamma_{0}$, satisfies
\begin{equation} \label{eq:conservation for gamma}
\int^{\pi}_{0}\alpha(\theta)\sin\theta d\theta=\beta \Gamma_{0}.
\end{equation}

After the change of variable $x=-\cos\theta$, the coupled system of evolution equations for $h(\theta,t)$ (\ref{eq:govern eqn for h}) and $\Gamma(\theta,t)$ (\ref{eq:governing eqn for gamma}) and their conservation properties (\ref{eq:conservation for h}) and (\ref{eq:conservation for gamma}) can be written more compactly as
\begin{equation}\label{eq:evolution eqn for h in x}
\frac{\partial h}{\partial t}+\frac{\partial}{\partial x} \left( h^2 (1-x^2) \left[Q_{1}h+\frac{1}{2}Q_{2} \right] \right)+2h\frac{\dot{R}(t)}{R(t)}=0\,,
\end{equation}
\begin{equation}\label{eq:evolution eqn for Gamma in x}
\frac{\partial \Gamma}{\partial t}+\frac{\partial}{\partial x}\left(h\Gamma (1-x^2) \left[ \frac{3}{2}Q_{1}h+Q_{2}\right] \right)=\frac{\partial}{\partial x}\left( \frac{D_s}{R^2(t)}(1-x^2)\frac{\partial \Gamma}{\partial x} \right)+\alpha-\Big( \beta +\frac{2\dot{R}}{R} \Big) \Gamma\,,
\end{equation}
and
\begin{equation}\label{eq:conservation eqn for h in x}
\frac{d}{dt}\left( R^2(t)\int^{1}_{-1}hdx\right)=0\,,
\end{equation}
\begin{equation}\label{eq:conservation eqn for Gamma in x}
\int^{1}_{-1}\alpha dx=\beta \Gamma_{0}\,,
\end{equation}
where
\begin{equation}
Q_{1}(h,\Gamma, x, t)=\frac{\rho g}{3\mu R(t)}+\frac{\sigma(\Gamma)}{3\mu R^{4} (t)}\frac{\partial}{\partial x}\left( 2h+\frac{\partial}{\partial x}\left((1-x^2)\frac{\partial h}{\partial x} \right)\right)\,,
\end{equation}
and
\begin{equation}
Q_{2}(\Gamma, x, t)=\frac{1}{\mu R^{2}(t)}\frac{\partial \sigma(\Gamma)}{\partial x}\,.
\end{equation}

% SUBSECTION

\subsection{Assumed forms of $\sigma(\Gamma)$ and $R(t)$} \label{subsection:gamma and sigma}

Equations (\ref{eq:evolution eqn for h in x}) and (\ref{eq:evolution eqn for Gamma in x}) provide a coupled system of equations involving $h(x,t)$ and $\Gamma(x,t)$. To close the system, one more relation that we need to specify is the dependence of $\sigma$ upon $\Gamma$. While a number of such relations have been stipulated and used previously, as mentioned in the Introduction, here we adopt a novel relation based on the observation that over a wide range of alveolar volumes, the pressure differential needed to to inflate the lungs remains relatively constant.
In order for the pressure jump across the alveolar fluid interface to remain nearly constant as the radius of the alveolus changes, surface tension must also change in order to keep the Laplace pressure constant. Consider an idealized state in the absence of gravity where the thin film coating the inside of the sphere modeling the alveolus is uniform and the surfactant distribution is also uniform. Assume that the Laplace pressure difference $\Delta p={2\sigma}/{R}$ is constant with respect to $R$ and see for what dependence $\sigma=\sigma(\Gamma(R))$ this would be possible. Since the surfactant is insoluble and the total amount of surfactant is conserved, we have $\Gamma \propto {1}/{R^2}$. Thus, only when $\sigma(\Gamma)={c_0}/{\sqrt{\Gamma}}$ where $c_0$ is a constant would the Laplace pressure remain constant. To obtain the constant $c_0$, we appeal to experimental measurements that show that when $\Gamma$ is at  its minimum, corresponding to the maximum alveolar radius, $\sigma(\Gamma)$ assumes its maximum value of approximately $\sigma_0=25$~mN/m.

To model the time-dependence of the radius of the alveoli, we appeal to the physiological parameter: percentage of total lung capacity (TLC). Let us assume that the radius of an alveolus is a periodic function of time with the form
\[\label{eq:radius wrt time}
R(t)=R_0(1-R_m+R_m \cos(\omega t))
\]
where $R_0>0$ the maximum radius, and $R_m$ the dimensionless oscillation amplitude, restricted to the range $0<R_m<0.5$. The alveolus has a radius of $R_0$ at peak inflation and one of $R_0(1-2R_m)$ at maximum exhalation. Total lung capacity is defined as the volume in the lungs at maximum inflation, so the fraction of total lung capacity at time $t$ (if all alveoli inflate and deflate in unison) is given by $({R(t)}/{R_0})^3$. Physiolgically, the percentage of TLC varies from 20\% to 100\% during spontaneous breathing; the approximation $R_m \approx 0.2$ puts us near the the lower end of that range.

% SUBSECTION

\subsection{Scaling and non-dimensionalization}

In this subsection we rewrite the evolution equations in scaled form and obtain the dimensionless groups that represent gravity, surface tension and Marangoni effects. Let $H$ be the average thickness of the thin film when the spherical alveolus reaches its maximum radius $R_0$. We list the physical parameters at that state in Table~\ref{tab:parameters}, some of which are taken from reference \cite{EMC2004}. We define dimensionless parameters $\hat{h}=h/H$ and $\hat{\Gamma}=\Gamma/\Gamma_{0}$ and define $\tau$ as an arbitrary time scale for the time being. The dimensionless equations take the forms
\begin{equation}\label{eq:dimensionless evolution eqn for h in x}
\frac{\partial \hat{h}}{\partial \hat{t}}+\frac{\partial}{\partial x} \left( \hat{h}^2 (1-x^2) \left[\hat{Q}_{1}\hat{h}+\frac{1}{2}\hat{Q}_{2} \right] \right)+2\hat{h}\frac{\dot{\hat{R}}}{\hat{R}}=0,
\end{equation}
\begin{multline}\label{eq: dimensionless evolution eqn for Gamma in x}
\frac{\partial \hat{\Gamma}}{\partial \hat{t}}+\frac{\partial}{\partial x}\left(\hat{h}\hat{\Gamma} (1-x^2) \left[ \frac{3}{2}\hat{Q}_{1}\hat{h}+\hat{Q}_{2}\right] \right)
\\=\frac{\partial}{\partial x}\left( \frac{\mathcal{D}}{\hat{R}^2}(1-x^2)\frac{\partial \hat{\Gamma}}{\partial x} \right)+\frac{\alpha\tau}{\Gamma_{0}}-\Big( \beta\tau +\frac{2\dot{\hat{R}}}{\hat{R}} \Big) \hat{\Gamma},
\end{multline}
where
\begin{equation}
\hat{Q}_{1}(\hat{h},\hat{\Gamma}, x, \hat{t})=\frac{\mathcal{G}}{\hat{R}}+\frac{\mathcal{S}\hat{\sigma}(\hat{\Gamma})}{\hat{R}^{4} }\frac{\partial}{\partial x}\left( 2h+\frac{\partial}{\partial x}\left((1-x^2)\frac{\partial \hat{h}}{\partial x} \right)\right),
\end{equation}
and
\begin{equation}
\hat{Q}_{2}(\hat{\Gamma}, x, \hat{t})=\frac{\mathcal{M}}{ \hat{R}^{2}}\frac{\partial \hat{\sigma}(\hat{\Gamma})}{\partial x}
\end{equation}
with
\begin{equation}\label{eq:scale for R and sigma}
\hat{R}(\hat{t})=\frac{R}{R_0}=1-R_m+R_m \cos(\omega \tau \hat{t}),\qquad \hat{\sigma}(\hat{\Gamma})=\frac{\sigma(\hat{\Gamma})}{\sigma_{0}}=(2\hat{\Gamma})^{-1/2}.
\end{equation}
The dimensionless relation between $\hat{\sigma}$ and $\hat{\Gamma}$ is based on the assumption outlined in subsection (\ref{subsection:gamma and sigma}) such that $\hat{\sigma}$ attains its maximum value of $1$ when a uniformly distributed $\hat{\Gamma}$ reaches its minimum value of $0.5$. Four dimensionless groups appear in these equations, definted by
\begin{equation}\label{eq:dimensionless groups}
\mathcal{G}=\frac{\tau \rho g H^2}{3\mu R_{0}},\quad \mathcal{S}=\frac{\tau \sigma_0 H^3}{3\mu R_{0}^{4}},\quad \mathcal{M}=\frac{\tau \sigma_{0} H }{\mu R_0^2},\quad \mathcal{D}=\frac{\tau D_s}{R_{0}^{2}}.
\end{equation}
These characterize gravity, surface tension, Marangoni effect and surface diffusion, respectively. In addition, the combination $\omega \tau$ represents the dimensionless frequency of the time-periodic breathing.

In Table~\ref{tab:parameters}, we collect the approximate experimental values of the parameters that relate to this system. We take the average radius of the alveoli to be approximately $0.1$~mm or 100 microns. The value of $\beta$ comes from assuming the half-life for the degradation of the surfactant to be about $5$ hours (while, in actuality, it may be as long as 24 hours). We calculate the dimensionless groups using experimental values. We see that the gravity parameter $\mathcal{G}$, the surface tension parameter $\mathcal{S}$ and the diffusion parameter $\mathcal{D}$ are comparable while the Marangoni number $\mathcal{M}$ is much larger than the other three groups. As we will see in the simulations, this causes the Marangoni effects to take place on a faster time scale, while the other effects determine the longer time dynamics.

For the rest of this paper, we will do simulations based on the equations (\ref{eq:dimensionless evolution eqn for h in x})--(\ref{eq:dimensionless groups}). We will drop the hats from the variables in the equation for notational convenience. We choose the time scale $\tau$ such that $\mathcal{S}=1$, which implies that $\tau\approx 12$~s, leading to $\mathcal{G}\approx 2.5$, $\mathcal{M}\approx 3\times 10^{4}$ and $\mathcal{D}\approx 1.2$. We also point out that $\beta \tau=6.72\times 10^{-4}\ll 1$, consistent with having a relatively long degradation time as compared to the time scale for gravitational drainage of the film or that for surface tension to act, and $\omega \tau=18.8$, which implies that the period of oscillations ($2\pi/\omega$) is about three times smaller than the time scale $\tau=12$~s.

\begin{table}
\caption{\label{tab:parameters} Values of the physical parameters when $R(t)=R_0$, the dimensionless groups, and some of their ratios.}
\centering
\begin{tabular}{|c|c|c|}

\hline\hline parameter & unit & value \\
\hline\hline  $\sigma_0$ & N/m & $2.5\times10^{-2}$   \\
\hline $g$ & m/s$^2$ & $9.8$ \\
\hline $\omega$ & 1/s & $\pi/2$ \\
\hline $\rho$ & kg/m$^3$ & $10^3$ \\
\hline $\mu$ & Pa$\cdot$s & $10^{-3}$ \\
\hline $R_0$ & m & $10^{-4}$ \\
\hline $H $ & m & $10^{-6}$ \\
\hline $D_{s}$ & m$^2/s$ & $10^{-9}$ \\
\hline $\beta$ & 1/s & $5.6\times 10^{-5}$ \\
\hline $\epsilon=H/L$ & 1 & $10^{-2}$ \\
\hline $\mathcal{G}={\tau \rho g H^2}/{(3\mu R_{0})}$ & $1$ & $ 3.3\times 10^{-2}\tau$ \\
\hline $\mathcal{S}={\tau \sigma_0 H^3}/{(3\mu R_{0}^{4})}$ & $1$ & $8.3\times 10^{-2}\tau$ \\
\hline $\mathcal{M}={\tau \sigma_{0} H }/{(\mu R_0^2)}$ & $1$ & $2.5\times 10^{3}\tau$ \\
\hline $\mathcal{D}={\tau D_s}/{R_{0}^{2}}$ & $1$ & $0.1\tau$ \\
\hline $ \mathcal{G}/\mathcal{S}$ & $1$ & $0.4$ \\
\hline $\mathcal{S}/\mathcal{M}$ & $1$ & $3.3\times10^{-5}$\\
\hline
\end{tabular}
%\label{tab:parameters}

\end{table}

% SECTION

\section{The Constant-Radius Case}

In this section, in order to understand the roles of surface tension, gravity and Marangoni effects without the complicating factor of changing radius, we study the model under the assumption that the radius of the alveolus is constant. All simulations are carried out using COMSOL Multiphysics. The next section considers the non-constant radius case where it is shown that through certain changes of variables, the equations can be simplified and analyzed without further detailed numerical simulations. Under the condition that $R(t)\equiv 1$, the simplified equations are
\begin{multline}
\frac{\partial h}{\partial t}+\frac{\partial}{\partial x}\left[h^{3}(1-x^2)\left(\mathcal{G}+\frac{\mathcal{S}}{\sqrt{2\Gamma}}\frac{\partial}{\partial x}\left(2h+\frac{\partial}{\partial x}\left((1-x^{2})\frac{\partial h}{\partial x}\right)\right)\right)\right. \\\left.-\frac{\mathcal{M}(1-x^2)}{2(2\Gamma)^{3/2}}\frac{\partial\Gamma}{\partial x}h^2\right]=0,
\label{eq:dimensionless h for const r}
\end{multline}
\begin{multline}
\frac{\partial\Gamma}{\partial t}+\frac{\partial}{\partial x}\left[\frac{3}{2}h^{2}\Gamma(1-x^2)\left(\mathcal{G}+\frac{\mathcal{S}}{\sqrt{2\Gamma}} \frac{\partial}{\partial x}\left(2h+\frac{\partial}{\partial x}\left((1-x^{2})\frac{\partial h}{\partial x}\right)\right)\right)\right.\\\left.-\frac{\mathcal{M} (1-x^2)}{(2\Gamma)^{3/2}} \frac{\partial\Gamma}{\partial x}h\Gamma\right]=\frac{\partial}{\partial x}\left(\mathcal{D}(1-x^{2})\frac{\partial\Gamma}{\partial x}\right)+\frac{\alpha\tau}{\Gamma_0}-\beta \tau \Gamma.
\label{eq:dimensionless gamma for const r}
\end{multline}

\begin{figure}
\hspace{-2cm}
\includegraphics[scale=0.42]{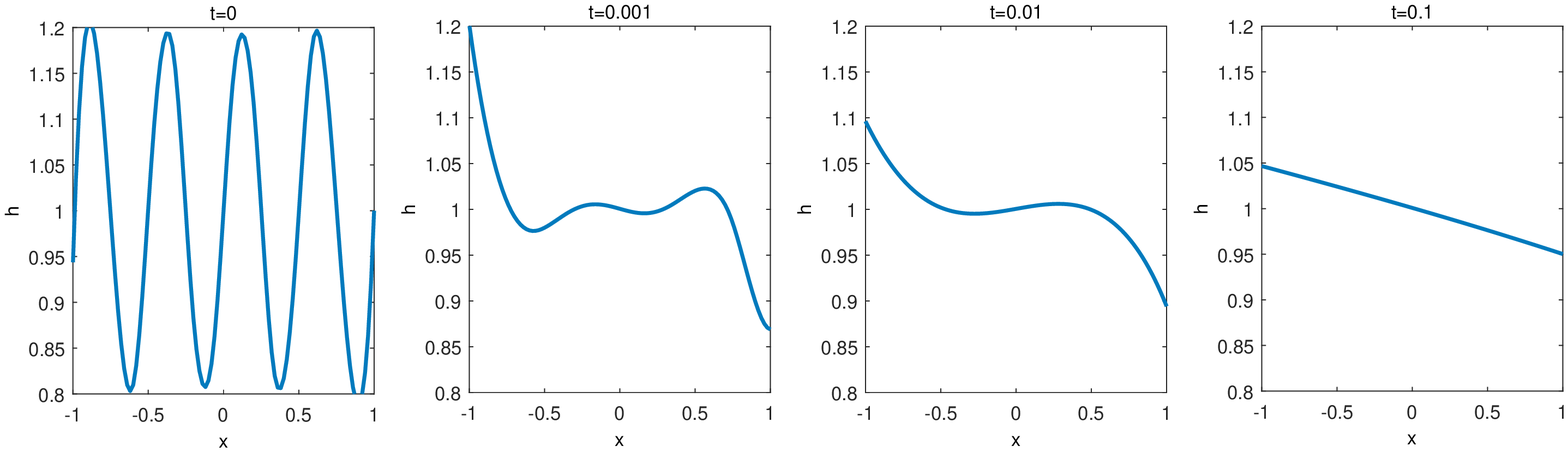}
\caption{$h(x,t)$ at times $t=$ 0, 0.001, 0.01, and 0.1 with in initially constant surfactant distribution. Surface tension and Marangoni effect are considered in the absence of gravity.}
\label{fig:1d no gravity}
\end{figure}

To start with, we only consider the surface tension and Marangoni effects. For the simulations, we thus take $\mathcal{G}=\alpha=\beta=0$ and $\mathcal{S}=1$, $\mathcal{M}=3\times10^4$ and $\mathcal{D}=1.2$. Fig.~\ref{fig:1d no gravity} shows the evolution of the film thickness starting with a sinusoidal perturbation but with an initially uniform surfactant distribution $\Gamma\equiv 0.5$. Surface tension drives the perturbed initial profile toward a state with uniform curvature and minimal area (i.e., an overall spherical shape) and the surfactant distribution will return to its constant state (not shown). We can observe from Fig.~\ref{fig:1d no gravity} that the final steady state of $h(x)$ is not a constant but a linear function of $x$. This is because in the absence of gravity, the final equilibrium state of the film, while having an overall spherical interface, need not have its center at the geometric center of the alveolus: the spherical alveolar fluid interface can be shifted slightly up or down (corresponding the film profile $h(\theta)=1+\varepsilon\cos\theta$ with a small $\epsilon$) and cause the steady state solution $h(x)$ not to be constant but to vary linearly in $x=-\cos\theta$. Depending on the initial perturbation in $h$, the inner spherical interface may end up shifted up or down relative to the outer spherical boundary of the alveolus.

\begin{figure}
\hspace{-2.cm}
\includegraphics[scale=0.42]{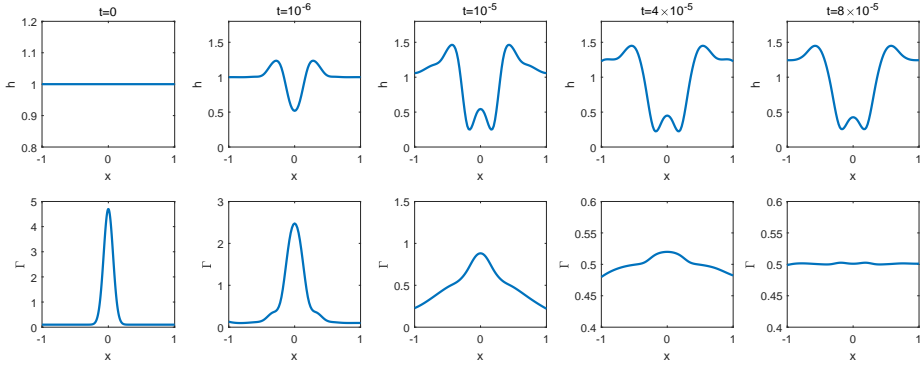}
\caption{$h(x,t)$ (top) and $\Gamma(x,t)$ (bottom)  at times $t=$ 0, 10$^{-6}$, 10$^{-5}$, $4\times 10^{-5}$ and $8\times 10^{-5}$ with a non-uniform initial surfactant distribution concentrated around the equator of the sphere. Surface tension and Marangoni effect are considered with no gravity. The short-time dynamics are seen in these plots.}
\label{fig:1d no gravity nonunif gamma short time}
\end{figure}

\begin{figure}
\hspace{-2cm}
\includegraphics[scale=0.42]{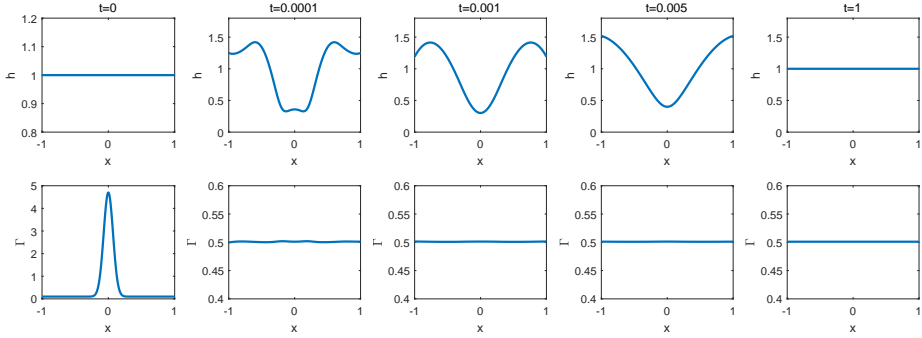}
\caption{$h(x,t)$ (top) and $\Gamma(x,t)$ (bottom) at times $t=$ 0, 0.0001, 0.001, 0.005 and 0.1 with a non-uniform initial surfactant distribution concentrated at the equator. Surface tension and Marangoni effect are included but with no gravity. The longer-time dynamics are evident in these plot.}
\label{fig:1d no gravity nonunif gamma}
\end{figure}

With a nonuniform initial $\Gamma$ distribution, surface tension and surface diffusion will still stabilize the system and drive $h$ and $\Gamma$ to uniform states as time tends to infinity, but the transient evolution is more complex. Figs.~\ref{fig:1d no gravity nonunif gamma short time} and \ref{fig:1d no gravity nonunif gamma} show the short- and long-time dynamics of the system, respectively, starting with an initial surfactant distribution that is highly concentrated near the equator of the spherical alveolus ($\theta=\pi/2$ or $x=0$). 
As seen in Fig.~\ref{fig:1d no gravity nonunif gamma short time}, the film thickness will exhibit some initial waviness for very short times: the high concentration of surfactant at the equator will cause the film to get thinner there very quickly as the Marangoni effect drives the film away from that low-tension area; meanwhile, due to Marangoni effects, the distribution of surfactant becomes nearly uniform very quickly during this same time. Over a longer time, as seen in Fig.~\ref{fig:1d no gravity nonunif gamma}, surface tension will drive $h$ toward a uniform distributions, while surface diffusion maintains the surfactant distribution near its uniform equilibrium state. To see the mathematical basis for the very rapid initial redistribution of surfactant, we can expand the $x$-derivative acting on the $\mathcal{M}$ term in equation (\ref{eq:dimensionless gamma for const r}) and combine it with the expanded term in the right-hand side. The coefficient of the diffusion term ${\partial ^2 \Gamma}/{\partial x^2}$ becomes $\mathcal{M}(2\Gamma)^{-3/2} (1-x^2) h\Gamma+\mathcal{D}(1-x^{2})$, in which the $\mathcal{M}$ term dominates the $\mathcal{D}$ term, having a magnitude of order of $10^4$. As such, the Marangoni term behaves as a diffusion effect and causes $\Gamma$ to become nearly constant on a diffusion time scale of $10^{-4}$. The relatively large Marangoni term thus causes any surfactant deposited on the alveolar fluid (e.g., by means of aerosolized drops) to quickly spread and cover the entire interface; while this can cause initial non-uniformities in the thickness of the alveolar fluid film, over a longer time, surface tension drives that back toward a uniform state.

\begin{figure}
\hspace{-2cm}
\includegraphics[scale=0.42]{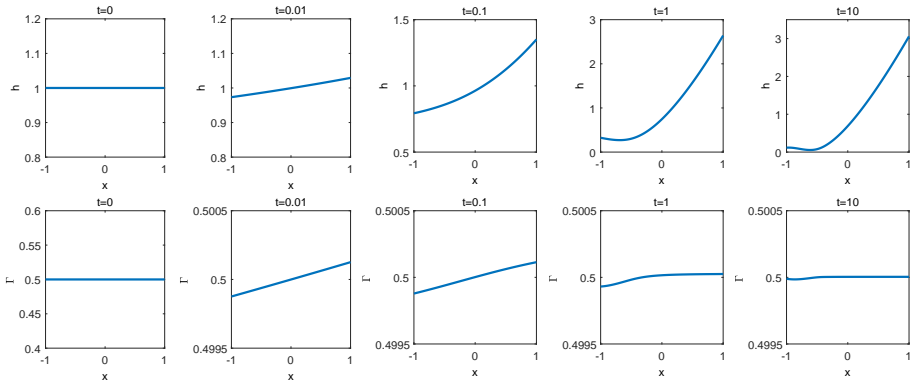}
\caption{$h(x,t)$ (top) and $\Gamma(x,t)$ (bottom) at times $t=$ 0, 0.01, 0.1, 1 and 10 with an initially uniform surfactant distribution. Gravity, surface tension and Marangoni effects are included. The film becomes thickest near the bottom due to gravitational drainage while surfactant concentration stays relatively constant.}
\label{fig:1d with gravity unif gamma}
\end{figure}

Although intuition might suggest that gravitational effects are negligible on the very small length scales of the alveolus and the thin film, the above scaling which leads to a parameter $\mathcal{G}$ of order unity, comparable to the surface tension effect $\mathcal{S}$, suggests that gravity does play some role even at the alveolus scale.
If we take gravity into consideration, Fig.~\ref{fig:1d with gravity unif gamma} shows that it will drive the thin film toward the bottom of the alveolus. As seen in the top row of that figure, after a dimensionless time of 10, the thickness profile is about three times larger than its starting value at the south pole (i.e., at $\theta=\pi$ or $x=1$). Interestingly, the profile is not monotonic and achieves its minimum not at the north pole, but slightly away from that. Since the Marangoni coefficient $\mathcal{M}$ is still large compared to the other coefficients, $\Gamma(x,t)$ remains nearly constant as seen in the bottom row of the figure (the vertical scale expands the immediate neighborhood of $\Gamma=0.5$).

\begin{figure}
\hspace{-2.cm}
\includegraphics[scale=0.42]{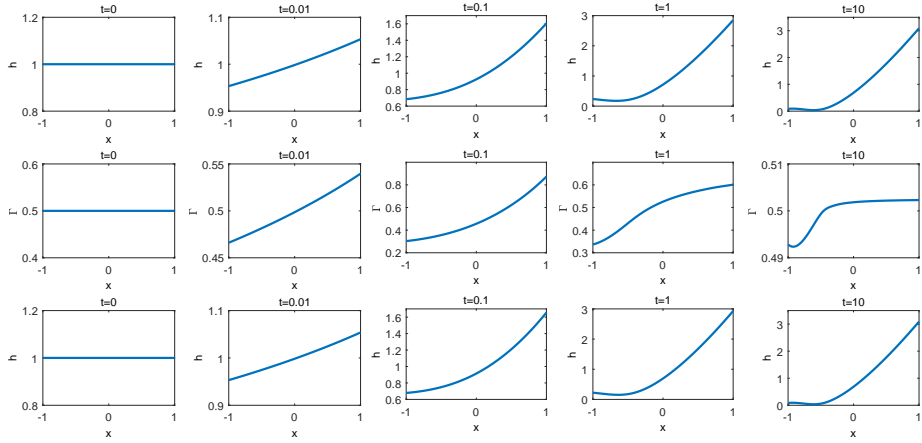}
\caption{$h(x,t)$ (top) and $\Gamma(x,t)$ (middle) for $\mathcal{M}=3$,  and $h(x,t)$ (bottom) without any surfactant, at times $t=$ 0, 0.01, 0.1,  1 and 10 with an initially uniform surfactant distribution. Gravity, surface tension and Marangoni effects are included. The evolution of the film is not affected much by the presence of surfactants; the film drains toward the bottom, and causes a transient redistribution of surfactants when present.}
\label{fig:1d with gravity unif gamma small m}
\end{figure}

In order to see what happens if the Marangoni parameter is not quite that large or when it is turned off completely, we compare the cases with a much smaller Marangoni parameter and with no Marangoni effect in Fig.~\ref{fig:1d with gravity unif gamma small m}. The top and bottom rows in the figure show the evolution of the film profile when $\mathcal{M}$ is 3 and 0, respectively. These appear fairly similar. The middle row shows the evolution of surfactant concentration $\Gamma$ for $\mathcal{M}=3$. In this case, surfactant concentration can deviate more from its equilibrium value of 0.5 and take longer to return to a uniform state as compared to the earlier results at the much higher value of the Marangoni parameter. 
%We can observe that the steady state for the cases with or without the consideration of Marangoni effect are nearly the same; meanwhile, Marangoni effect can slow down the dynamics towards the equilibrium.

\begin{figure}
\hspace{-2.cm}
\includegraphics[scale=0.42]{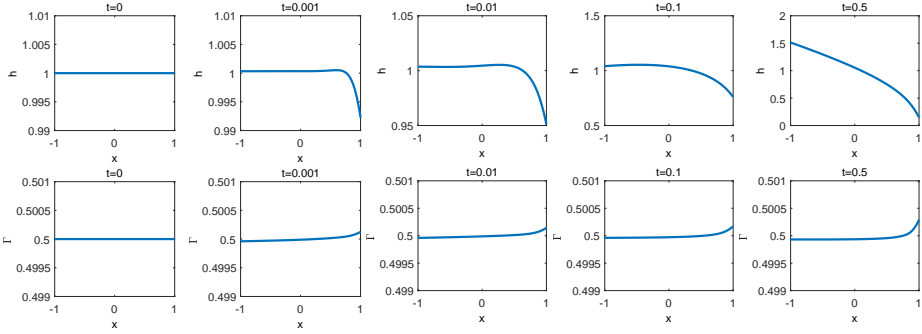}
\caption{$h(x,t)$ (top) and $\Gamma(x,t)$ (bottom) at times $t=$ 0, 0.001, 0.01, 0.1 and 0.5 with a source concentrated at the bottom: $x=1$. Surface tension and Marangoni effects are included, but gravity is absent. The presence of the source causes the film to become thinnest at the source and thickest at the opposite pole. Surfactant concentration is slightly higher near the source but fairly uniform away from that.}
\label{fig:no gravity big source const R}
\end{figure}

At this point we also add the surfactant production and degradation terms to the analysis. In order to observe the effects caused by these more clearly, we perform the simulations in the absence of gravity. After scaling, in order to make the total amount of surfactant constant, we need $\int^{1}_{-1}({\alpha}/{\Gamma_0})dx=\beta$. We consider a strong source of surfactant that is concentrated at the bottom of the sphere. In this case, extra surfactant concentration will cause lower surface tension at the bottom and lead to an upward flow along the walls of the alveolus. We do the simulation using $\beta=1$ and ${\alpha (x)}/{\Gamma_0}=11.284\times\exp[-100(x-1)^2]$ and the result is shown in Fig.~\ref{fig:no gravity big source const R}. In the top row, we observe that the film initially thins out at the south pole (near $x=1$) and gradually assumes a monotonic profile which is thickest at the north pole away from the source and thinnest at the bottom where the source is located. The surfactant concentration, shown in the bottom rows, shows a slight maximum near the source but is relatively uniform away from it. This suggests that if Type 2 cells in the alveoli are the sources of pulmonary surfactant, the alveolar fluid layer may be thinnest just above those Type 2 cells.

Another useful approach which could provide insight into the behavior of alveolar fluid in a microgravity environment is to perform a regular perturbation analysis when the gravity parameter $\mathcal{G}$ is very small but nonzero. Consider equation (\ref{eq:dimensionless h for const r}) in the presence of surface tension in a microgravity environment, i.e., when $\mathcal{G}=\delta\ll1$, without considering the effects of production or degradation of surfactant. Assume that the solution $h$ has an asymptotic expansion of the form:
\begin{equation}
h=h_0+\delta h_1+\delta^2 h_2 +\cdots
\end{equation}
The equation that describes the leading order term $h_0$ is found to be 
\begin{equation}
\frac{\partial h_0}{\partial t}+\frac{\partial}{\partial x}\left(h_{0}^{3}(1-x^2)\mathcal{S}\frac{\partial}{\partial x}\left(2h_0+\frac{\partial}{\partial x}\left((1-x^2)\frac{\partial h_0}{\partial x} \right) \right) \right)=0,
\end{equation}
whose steady state solution can be the constant $h_{0}=1$, depending on the initial condition. At the first order, the equation becomes
\begin{multline*}
\frac{\partial h_{1}}{\partial t}+\frac{\partial}{\partial x}\left(h_{0}^{3}(1-x^2)+(1-x^2)\mathcal{S}\left( h_{0}^{3}\frac{\partial}{\partial x}\left(2h_1+\frac{\partial}{\partial x}\left((1-x^{2})\frac{\partial h_1}{\partial x}\right)\right) \right.\right. \\
\left. \left. +3h_{0}^{2}h_{1}\frac{\partial}{\partial x}\left(2h_0+\frac{\partial}{\partial x}\left((1-x^{2})\frac{\partial h_0}{\partial x}\right)\right) \right)\right)=0\,,
\end{multline*}
whose steady state solution corresponding to the state $h_0=1$ satisfies
\begin{equation}
\frac{\partial}{\partial x}\left((1-x^2)+(1-x^2)\mathcal{S}\frac{\partial}{\partial x}\left(2h_1+\frac{\partial}{\partial x}\left((1-x^{2})\frac{\partial h_1}{\partial x}\right) \right)\right)=0.
\end{equation}
The general solution of this fourth-order equation can be written as
\begin{multline*}
h_1=C_1+C_{2}x+\frac{1}{12\mathcal{S}}[-4x-\ln(1-x)(3-2x+3\mathcal{S}(C_{3}x+C_4))
\\ 
+\ln(1+x)((2x+3)+3\mathcal{S}(C_{3}x+C_4)) ].
\end{multline*}

%\marginpar{Which figures are these last paragraphs referring to or are we not including them anymore? I'll edit these if we do include the figures.}
%If we use the parameters comes from the experiments and an uniform distributed source (${\alpha (x)}/{\Gamma_0}=\beta\sin(2\pi x)^2$), the simulation results are shown in Fig.
%

%Next we add source into this system. We firstly assume  that there is a source at the bottom and sink at the top. In this case we choose $\alpha(x)=e^{-10(x-1)^2}-e^{-10(x+1)^2}$ and $beta=0$. Fig. \ref{fig:no g with source} shows that in the absence of gravity the steady state of surfactant concentration is not a constant any more and the shape of thin film stays as a perfect sphere but touches the spherical substrate at the bottom. In the presence of gravity, there is a compete between gravity and Marangoni effect. Fig. \ref{fig:with g with source} shows the convergence to steady state of this system. There is an internal flow in this case.
%

% SECTION

\section{Non-constant Radius Case}

% SUBSECTION
% \subsection{Simulation results}

In this section we consider the case where the radius of the spherical alveolus changes with respect to time periodically. We take the scaled radius to be given by equation (\ref{eq:scale for R and sigma}), with the full evolution equations given in Eqs.~(\ref{eq:dimensionless evolution eqn for h in x})--(\ref{eq:scale for R and sigma}). We run the simulations using the experimental parameters from Table~\ref{tab:parameters}, except we choose a moderate Marangoni number $\mathcal{M}=3$ in order not to overwhelm the other effects by the fast acting Marangoni term.

\begin{figure}
\hspace{-2cm}
\includegraphics[scale=0.42]{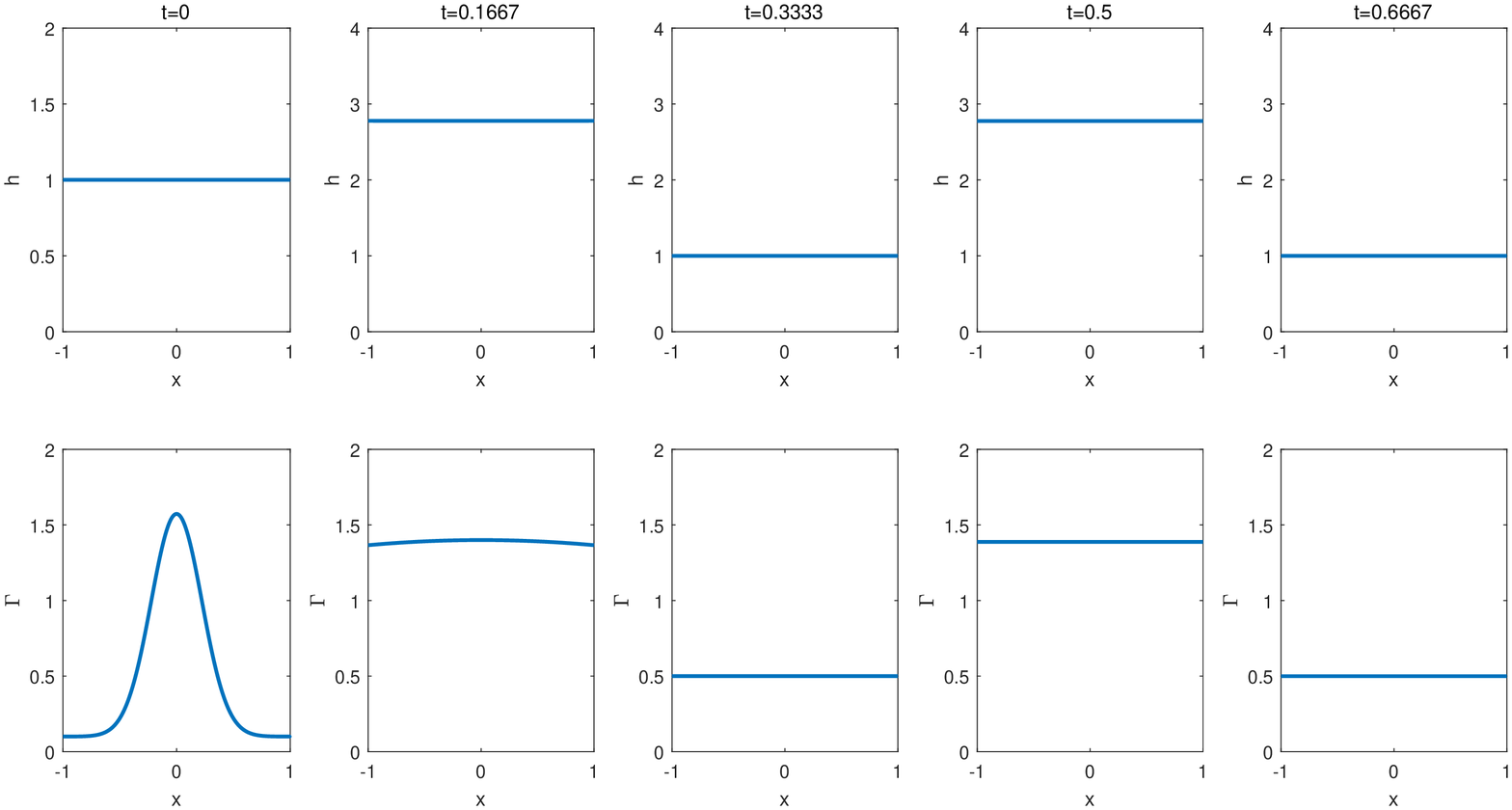}
\caption{$h(x,t)$ (top) and $\Gamma(x,t)$ (bottom) at times $t=$ 0, 0.1667, 0.3333, 0.5 and 0.6667.  The sphere radius changes periodically with time. Surface tension and Marangoni effects are included, but gravity is absent. The first, third and fifth column are when the sphere is at its maximum radius, and second and fourth column are when the sphere is at its minimum radius.}
\label{fig:nonconst R no g long time}
\end{figure}

\begin{figure}
\hspace{-2.cm}
\includegraphics[scale=0.42]{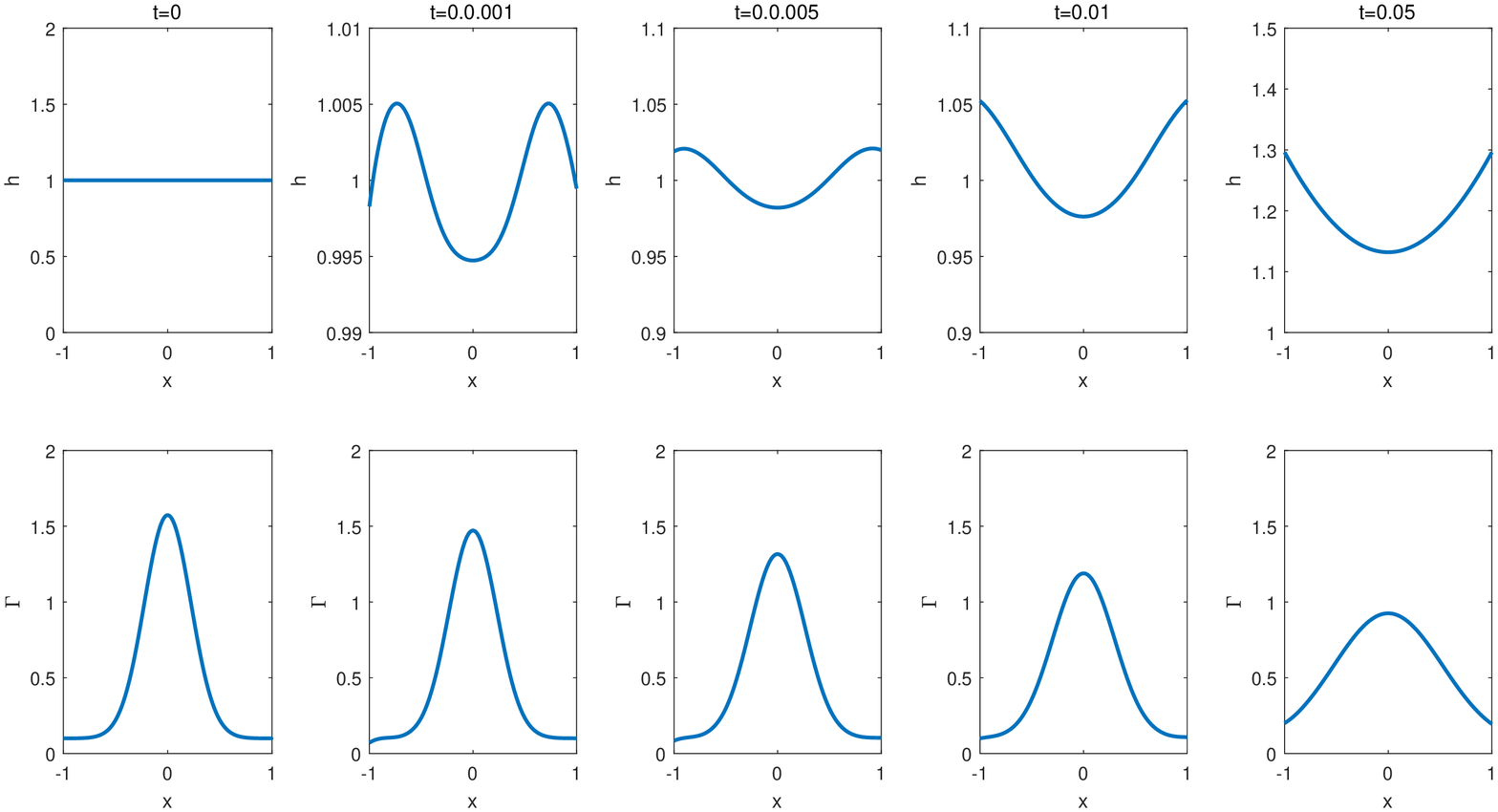}
\caption{$h(x,t)$ (top) and $\Gamma(x,t)$ (bottom) at times $t=$ 0, 0.001, 0.005, 0.01 and 0.05. The sphere radius changes periodically with time. Surface tension and Marangoni effects are included, but gravity is absent.}
\label{fig:nonconst R no g short time}
\end{figure}

We first study the case in the absence of gravity. Long-time and short-time behaviors are shown in Fig.~\ref{fig:nonconst R no g long time} and Fig.~\ref{fig:nonconst R no g short time} respectively. The period of oscillations is ${1}/{3}$ and the time difference between adjacent  columns in Fig.~\ref{fig:nonconst R no g long time} is half of a period. We see from the figure that both $h$ and $\Gamma$ are nearly uniform at later times, but their values change due to the change of radius, in order to conserve mass. The first, third and fifth columns correspond to times when the radius of the alveolus is at its maximum, at which times the values of $h$ and $\Gamma$ reach their minimum, while the second and fourth columns are the opposite. For short times, Fig.~\ref{fig:nonconst R no g short time} shows the result within half of a period. We see that the results are similar to Fig.~\ref{fig:1d no gravity nonunif gamma short time}, but the average values of $h$ and $\Gamma$ are increasing due to the initial decrease in the radius of the alveolus. 

\begin{figure}
\hspace{-2.cm}
\includegraphics[scale=0.42]{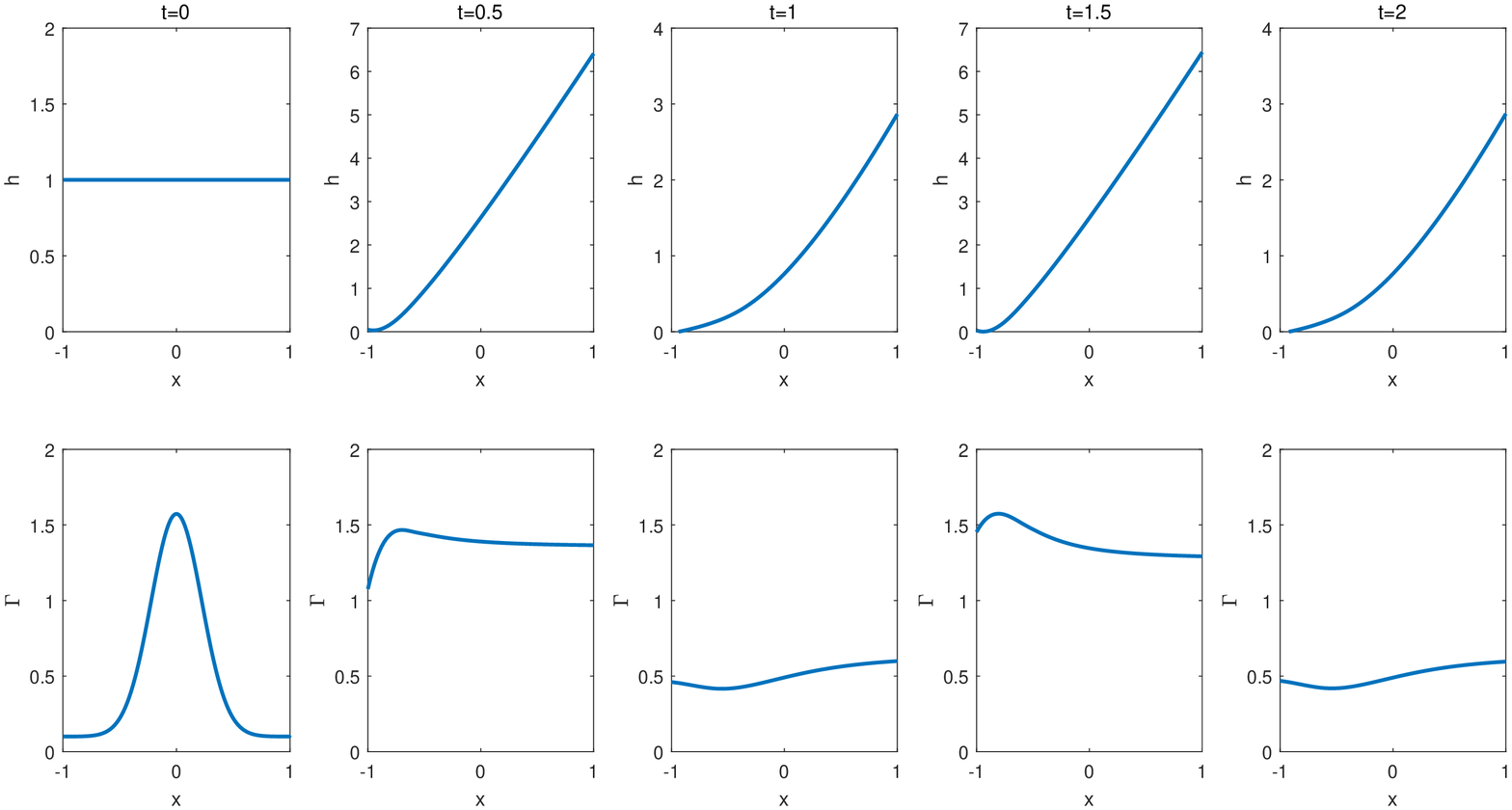}
\caption{$h(x,t)$ (top) and $\Gamma(x,t)$ (bottom) at times $t=$ 0, 0.5, 1, 1.5 and 2. The sphere radius changes periodically with time. Gravity, surface tension and Marangoni effects are included.}
\label{fig:nonconst R with g long time}
\end{figure}

\begin{figure}
\hspace{-2.cm}
\includegraphics[scale=0.42]{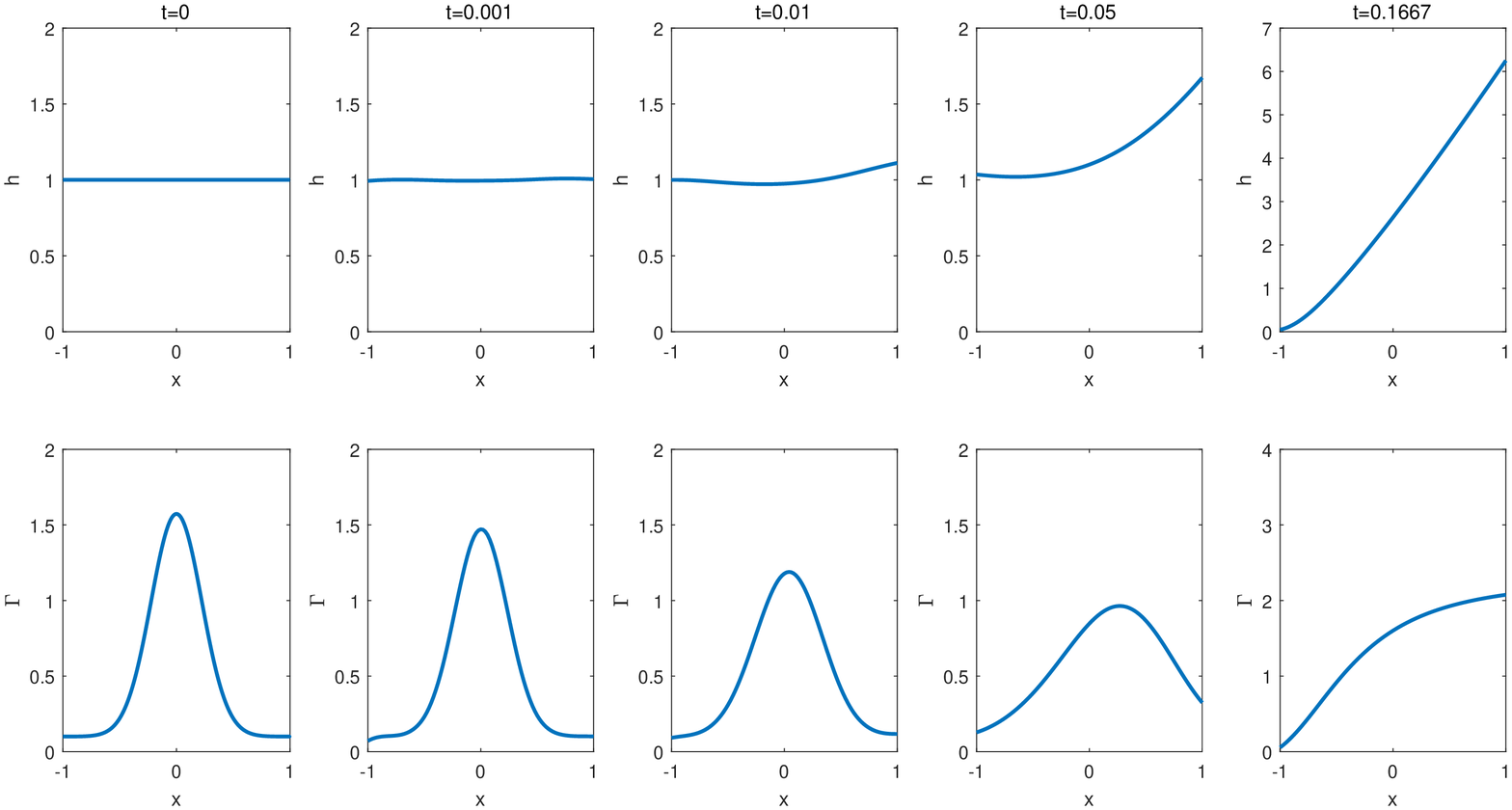}
\caption{$h(x,t)$ (top) and $\Gamma(x,t)$ (bottom) at times $t=$ 0, 0.001, 0.01, 0.05 and 0.1667. The sphere radius changes periodically with time. Gravity, surface tension and Marangoni effects are included.}
\label{fig:nonconst R with g short time}
\end{figure}

In the presence of gravity, long-time and short-time simulations are shown in Fig.~\ref{fig:nonconst R with g long time} and Fig.~\ref{fig:nonconst R with g short time}. A dry zone appears near the top due to gravity when time gets large. Fig.~\ref{fig:nonconst R with g long time} also shows that the monotonicity of $\Gamma$ changes when the radius of the sphere reaches its maximum and minimum values.

% SUBSECTION

\subsection{Scaling with respect to $R(t)$}

The governing equations that describe $h(x,t)$ and $\Gamma(x,t)$ are shown in equations (\ref{eq:dimensionless evolution eqn for h in x}) and (\ref{eq: dimensionless evolution eqn for Gamma in x}). Multiplying both sides of both equations by $R^{2}(t)$ and combining terms allow us to write
\begin{equation}
\frac{\partial }{\partial t}\ln(R^{2}(t)h)+\frac{\partial}{\partial x} \left(R^{2}(t) h^2 (1-x^2) \left[Q_{1}h+\frac{1}{2}Q_{2} \right] \right)=0\,,
\end{equation}
\begin{multline}
\frac{\partial}{\partial t}\ln(R^{2}(t)\Gamma)+\frac{\partial}{\partial x}\left(R^{2}(t) h\Gamma (1-x^2) \left[ \frac{3}{2}Q_{1}h+Q_{2}\right] \right)\\=\frac{\partial}{\partial x}\left( \mathcal{D}(1-x^2)\frac{\partial \Gamma}{\partial x} \right)+\frac{\alpha\tau R^{2}(t)}{\Gamma_0}-\beta\tau R^{2}(t)\Gamma\,.
\end{multline}
With the change of dependent variables $\tilde{h}(x,t)=h(x,t) R^2(t)$ and $\tilde{\Gamma}(x,t)=\Gamma(x,t)R^{2}(t)$, we can rewrite these as
\begin{multline}\label{eq:equation for h with nonconst R after scaling}
\frac{\partial \tilde{h}}{\partial t}+\frac{\partial}{\partial x}\left(\tilde{h}^2 (1-x^2)\left(\left[\frac{\mathcal{G}}{R^{5}(t)}+ \frac{\mathcal{S}}{\sqrt{2\tilde{\Gamma}}R^{9}(t)}  \frac{\partial}{\partial x}\left( 2\tilde{h} +\frac{\partial}{\partial x}\left( (1-x^2)\frac{\partial \tilde{h}}{\partial x}\right)  \right)\right]\tilde{h}\right.\right. \\\left.\left.-\frac{(2\tilde{\Gamma})^{-3/2}\mathcal{M}}{2R(t)}\frac{\partial \tilde{\Gamma}}{\partial x}\right) \right)
=0\,,
\end{multline}
and
\begin{multline}
\label{eq:equation for gamma with nonconst R after scaling}
\frac{\partial \tilde{\Gamma}}{\partial t}+\frac{\partial}{\partial x}\left(\tilde{h}\tilde{\Gamma} (1-x^2)\left(\frac{3}{2}\left[\frac{\mathcal{G}}{R^{5}(t)}+ \frac{\mathcal{S}}{\sqrt{2\tilde{\Gamma}}R^{9}(t)}  \frac{\partial}{\partial x}\left( 2\tilde{h} +\frac{\partial}{\partial x}\left( (1-x^2)\frac{\partial \tilde{h}}{\partial x}\right)  \right)\right]\tilde{h}\right.\right.\\\left.\left.-\frac{(2\tilde{\Gamma})^{-3/2}\mathcal{M}}{R(t)}\frac{\partial \tilde{\Gamma}}{\partial x}\right) \right)
=\frac{\partial}{\partial x}\left( \frac{\mathcal{D}}{R^{2}(t)}(1-x^2)\frac{\partial \tilde{\Gamma}}{\partial x} \right)+\frac{\alpha\tau R^{2}(t)}{\Gamma_0}-\beta\tau \tilde{\Gamma}.
\end{multline}
We observe that these equations have the same forms as the constant-radius case, only with parameters that depend on various powers of $R(t)$. It should be noted that as $R(t)$ varies between its maximum scaled value of unity and the lower bound $1-2R_m\approx 0.6$ (when $R_m=0.2$ as suggested earlier), factors such as $R^5$ and $R^9$ attain much smaller minimum values and change the relative importance of the effects of gravity, surface tension, Marangoni flow, surface diffusion and surfactant production rate.
%\marginpar{Should we not include any simulation results with a periodic R(t) in this section?}

When some of those effects act in isolation, a mathematical trick allows us to remove the time dependent radius $R(t)$ from the equation by defining a new time-like variable. For instance, if we consider the problem under the effect of surface tension only, we can rewrite Eq.~(\ref{eq:equation for h with nonconst R after scaling}) after defining a new time-like variable $\tilde{t}=\int_{0}^{t}R^{-9}(\tau)d\tau$, in the form
\begin{equation}\label{eq:eqn h with nonconst R without g}
\frac{\partial \tilde{h}}{\partial \tilde{t}}+\frac{\partial}{\partial x}\left(\tilde{h}^2 (1-x^2) \mathcal{S}\sigma(\tilde{\Gamma})  \frac{\partial}{\partial x}\left( 2\tilde{h} +\frac{\partial}{\partial x}\left( (1-x^2)\frac{\partial \tilde{h}}{\partial x}\right) \tilde{h} \right)\right)=0.
\end{equation}
This equation admits $\tilde{h}\equiv 1$ as its attracting steady state solution, which implies that $h(t)={1}/{R^{2}(t)}$ is a global attractor for equation (\ref{eq:dimensionless evolution eqn for h in x}) when only surface tension is acting; that is, non-uniform initial film profiles tend to the uniform state that depends only upon time and varies as the radius changes while conserving the fluid volume.

% SUBSECTION

\subsection{Effect of Gravity}

One mathematically interesting case which can be analyzed readily using the method of characteristics is when gravity alone acts on the film, while surface tension and all surfactant effects are absent (i.e., only parameter $\mathcal{G}$ is nonzero). Upon defining the new time-like variable $\tilde{t}=\mathcal{G}\int_{0}^{t}R^{-5}(\tau)d\tau$, equation (\ref{eq:equation for h with nonconst R after scaling}) can be written as
\begin{equation}
\frac{\partial\tilde{h}}{\partial\tilde{t}}+\frac{\partial}{\partial x}(\tilde{h}^{3}(1-x^2))=0\,,
\end{equation}
having the conservation form. Its expanded form reads
\begin{equation}
\frac{\partial \tilde{h}}{\partial \tilde{t}}+3\tilde{h}^2(1-x^2)\frac{\partial \tilde{h}}{\partial x}=2x\tilde{h}^3\,,
\end{equation}
which is a first-order quasilinear PDE amenable to the method of characteristics. The characteristic form of this equation is equivalent to a system of two first-order nonlinear ODEs:
\begin{equation}\label{eq:characteristic eqn}
\frac{d\tilde{h}}{d\tilde{t}}=2x\tilde{h}^3 \quad \mbox{along paths:}\quad
\frac{dx}{d\tilde{t}}=3(1-x^2)\tilde{h}^2\,. 
\end{equation}
%\marginpar{Somehow, I think we should take this solution one step further, past (4.11).}
With a change of variable $y=\tilde{h}^{-2}$, we can rewrite the previous two equations as
\begin{equation}
\frac{dx}{d\tilde{t}}=\frac{3(1-x^2)}{y}\,,\quad \frac{dy}{d\tilde{t}}=-4x\,.
\end{equation}
Upon dividing these we find
\begin{equation}
\frac{dy}{dx}=-\frac{4xy}{3(1-x^2)}\,,
\end{equation}
whose solution in terms of $\tilde{h}$ is given by
\begin{equation}
\label{constancy}
\tilde{h}(\tilde{t})\,(1-x(\tilde{t})^2)^{1/3}=\mbox{constant}\,.
\end{equation}
Although the characteristics obtained from equations (\ref{eq:characteristic eqn}) cannot be obtained in closed form, we can make progress at the two poles. Notice that ${dx}/{dt}=0$ at $x=\pm1$, which shows that characteristics through $x=\pm1$ are perpendicular to the $x$-axis. As such, we can find $h(x=\pm1,t)$ by solving the first equation in (\ref{eq:characteristic eqn}):
\[
\frac{d\tilde{h}(\pm1,t)}{dt}=\pm 2\tilde{h}^{3},\quad \tilde{h}(x,0)=1\,,
\]
assuming a uniform initial profile with $h=1$ throughout. This yields $\tilde{h}(\pm1,t)=1/{\sqrt{1\mp 4t}}$, which shows that at the north pole, the film thickness tends to zero for long times at the rate $t^{-1/2}$, and at the south pole, the film thickness will blow up in  finite time at the point $t=0.25$. Since surface tension is a stabilizing effect, the rate of convergence to zero at the north pole when both gravity and surface tension are present will be slower than $t^{-1/2}$.

We can also integrate the characteristic equations (\ref{eq:characteristic eqn}) numerically starting at various initial points $-1<x_o<1$. While the results have not been displayed here, we find that all characteristics not emanating from the north pole ultimately reach the south pole in finite time, at which time the solution diverges. The closer $x_o$ is to the south pole, the faster the blow-up occurs, with the blow-up time approaching 0.25 as $x_o$ tends to 1. So, up to the first blow-up time of $t=0.25$ at the south pole itself, the solution remains regular everywhere else and the numerical solution does indeed satisfy condition (\ref{constancy}) along the characteristics. It should be noted that this analysis applies equally to the case where the radius of the alveolus is a constant. The introduction of the time-like variable $\tilde{t}$ removes any time-dependence of radius $R$ from the analysis.
%\marginpar{I have not yet edited this entire subsection. In (4.13), isn't R a function of t? So, that can't be a steady state solution for h1. Should we include this anyway?}

% DISCUSSION

\section{Discussion and Further Research Questions}

One key observation from our numerical simulations with parameters chosen based on real experimental values is that if a small amount of surfactant is delivered into the alveolus in the form of an aerosol drop and makes contact with the alveolar fluid layer at a point, the time it takes for it to spread uniformly over the entire interfacial area to form a monolayer is quite fast ($\approx$ 10$^{-3}$ sec) but the required time for the thin film of alveolar fluid to regain a relatively uniform thickness after the surfactant droplet deposition could be much longer ($\approx$ 1~min).

We derived our model under the assumption that lung surfactant forms a monolayer inside the alveolar compartment but the real structure is much more complicated. It was experimentally observed that the microscopic membrane formed by lung surfactant consists of multiple layers and can even form membrane tubes and other myelin figures \cite{Elisa}. These stacked interfaces with microscopic tubes and helices form a viscoelastic bulk material with near zero surface tension which has not been studied fully and is not yet well understood. The authors of \cite{Elisa} came to the conclusion that lung surfactant could be modeled as a highly viscous multilayered material that supports any Laplace pressure simply by viscoelastic shear of the tightly stacked multilamellar membranes.

Also, our model did not take into account connections between multiple alveoli and the role of surfactant in the averaging of alveoli sizes. Surfactant helps all alveoli in the lung expand at nearly the same rate, since if one expands more quickly, it will experience a rise in surface tension slowing its rate of expansion. It also means the rate of shrinking is more uniform, since if one reduces in size more quickly, its surface tension will reduce more, so the other alveoli can contract more easily and catch up to it. Surfactant reduces surface tension more readily when the alveoli are smaller because the surfactant is more concentrated. We plan to capture this size regulation effect of lung surfactant in our future research by modeling a multi-alveoli system.
 
Synthetic pulmonary surfactants, which are currently administrated to prematurely born infants, differ in surface viscosity. Viscosity is believed to influence the rate, extent and uniformity of distribution of surfactant in the lungs. Surfactants with lower surface viscosity are preferred in neonatology for endotracheal application because they allow a more uniform and rapid distribution of the instilled surfactant with less loss due to coating of the upper airways. Despite its widespread use, the optimal method of surfactant administration in prematurely born infants has not been clearly determined yet \cite{nina}. Several aspects of administration techniques that can influence surfactant delivery into the pulmonary airways including the bolus volume, injection rate, gravity and orientation, ventilation strategies, alveolar recruitment, and viscosity and surface tension of the fluid instilled are discussed in the review article \cite{nina}. 
 
An additional challenging topic would be to model the entire dynamics of synthetic surfactant delivery into the alveoli. From the bronchi, the dividing airways become progressively smaller with an estimated $23$ divisions before ending at an alveolus. High-resolution MRI scans of the lungs and airways allow one to visualize only the first $8$ branches and, of these, only the first $6$ can be reconstructed from the MRI image by 3D printing. As such a fluid dynamic model is ideal for optimizing the parameters for the process of surfactant delivery into the lungs.

\bigskip

\noindent{\bf Acknowledgement:} This work was partially supported by a grant from the Simons Foundation (\#275088 to Marina Chugunova).

%\begin{acknowledgements}
%\end{acknowledgements}


\begin{thebibliography}{10}

%\bibitem{Alt}
%\textcolor{red}{H.W. Alt. Lineare Funktionalanalysis. Springer-Verlag, Berlin, 1985.}
%\bibitem{type2}
%Andreeva A.V., Kutuzov M.A., Voyno-Yasenetskaya T.A.:
%\newblock{Regulation of surfactant secretion in alveolar type II cells,}
%\newblock{American Journal of Physiology - Lung Cellular and Molecular Physiology, Vol. 293 no. 2, L259-L271} (2007)

%\bibitem{Bar1}
%Barrett J.W., Garcke H., N\"{u}rnberg R.:
%\newblock Finite Element Approximation of Surfactant Spreading on a Thin Film,
%\newblock {SIAM J. Numer. Anal.}, 41(4):1427--1464 (2004)

%\bibitem{Bar2}
%Barrett J.W., N\"{u}rnberg R.:
%\newblock Convergence of a finite-element approximation of surfactant
%spreading on a thin film in the presence of van der Waals forces,
%\newblock { IMA Journal of Numerical Analysis}, 24(2):323--363 (2004)


%\bibitem{B8}
%Bernis F., Friedman A.:
%\newblock Higher order nonlinear degenerate parabolic equations,
%\newblock { J. Differential Equations}, 83(1):179--206 (1990)
%
%
%\bibitem{Bert1998}
%Bertozzi A.L., Pugh M.C.:
%\newblock Long-wave instabilities and saturation in thin film equations,
%\newblock {Comm. Pur. Appl. Math.}, LI:625--651 (1998)

%\bibitem{Bert2000}
%Bertozzi A.L., Pugh M.C.:
%\newblock  Finite-time blow-up of solutions of some long-wave unstable thin film equations,
%\newblock {Indiana Univ. Mathematics J.}, 49(4):1323--1366 (2000)

\bibitem{BG}
Borgas M.S., Grotberg J.B.:
\newblock Monolayer flow on a thin film.
\newblock { J. Fluid Mech.}, 193:151--170 (1988)

\bibitem{space2}
Bryndina I., Vasilieva N.: 
\newblock{Effects of simulated microgravity on surfactant and water balance of lung in animals with different resistance to stress,}
\newblock{40th COSPAR Scientific Assembly, Vol 40} (2014)







%\bibitem{BM}
%Ha\"{\i}m Brezis, Moshe Marcus.
%\newblock Hardy's inequalities revisited.
%\newblock {\em Annali della Scuola Normale
%Superiore di Pisa - Classe di Scienze, Ser. 4.}, 25
%(1-2):217--237, 1997.



\bibitem{Carnielli}
Carnielli V.P., Zimmermann L.J., Hamvas A., and Cogo P.E.: 
\newblock{Pulmonary surfactant kinetics of the newborn infant: novel insights from studies with stable isotopes,}
\newblock{Journal of Perinatology, 29:S29 - S37 (2009)}

\bibitem{castranova1988}
Castranova V, Rabovsky J, Tucker J.H., Miles P.R.: 
\newblock{The alveolar type II epithelial cell: a multifunctional pneumocyte,}
\newblock{Toxicol Appl Pharmacol,93(3):472-83 (1988)}


%\bibitem{CK}
%Chou K.S., Kwong Y.C.: 
%\newblock Finite time rupture for thin films under
%van der Waals forces,
%\newblock { Nonlinearity}, 20(2):299?-317 (2007)
%
%
%\bibitem{Chu2010}
%Chugunova M., Pugh  M.C., Taranets  R.: 
%\newblock Nonnegative solutions for a long-wave unstable thin film equation with convection,
%\newblock { SIAM J. Math. Anal.}, 42(4):1826--1853 (2010)

\bibitem{surf1}
Chugunova M., Taranets R.: 
\newblock{The interface dynamics of a surfactant drop on a thin viscous film,}
\newblock { European Journal of Applied Mathematics, doi.org/10.1017/S0956792516000474, 1 - 31 (2016)}

\bibitem{surf2}
Chugunova M., Taranets R.
\newblock{ Nonnegative weak solutions for a degenerate system modelling the spreading of surfactant on thin films,}
\newblock{ Applied Mathematics Research Express, 1, 102 - 126 (2013)}

\bibitem{Crapo}
Crapo J.D., Young S.L., Fram E.K., Pinkerton K.E., Barry B.E., and Crapo R.O.:
\newblock{Morphometric Characteristics of Cells in the Alveolar Region of Mammalian Lungs,}
\newblock{American Review of Respiratory Disease, 128(2P2):S42 - S46 (1983)}

%\bibitem{Dar2010}
%Dargaville P.A., Aiyappan A., Cornelius A., Williams C., DePaoli A.G.: 
%\newblock Preliminary evaluation of a new technique of minimally invasive surfactant therapy, 
%\newblock {Arch. Dis. Child. Fetal Neonatal. Ed.} (2010)


%\bibitem{deWitGallez}
%De Wit A., Gallez D., Christov C.I.: 
%\newblock Nonlinear evolution equations for thin liquid films with insoluble surfactant, 
%\newblock {Phys.Fluids}, 6:3256-3266 (1994)
%
%\bibitem{DiB-0} DiBenedetto E., Friedman A.: 
%\newblock   H\"{o}lder estimates for nonlinear degenerate
%parabolic systems,
%\newblock{J. Reine Angew. Math.}, 363: 217-220 (1985)

%\bibitem{DiB}
%DiBenedetto E.:
%\newblock Degenerate parabolic equations,
%\newblock {Universitext. Springer-Verlag, New York}, xvi+387 pp, 1993 (2012)

%\bibitem{Ed}
%{\`E}{\u\i}del'man S.D.:
%\newblock {Parabolic systems},
%  North-Holland Publishing Co., Amsterdam (1969)
  
\bibitem{EMC2004}
Edmonstone B.D., Matar O.K., and Craster R.V.:  Flow of surfactant-laden thin films down an inclined plane, Journal of engineering mathematics 50.2: 141-156 (2004)

\bibitem{EHLW}
Escher J., Hillairet M., Lauren\c{c}ot Ph., Walker Ch.:
\newblock Global weak solutions for a degenerate parabolic system modeling the spreading of insoluble surfactant,
\newblock to appear in {Indiana Math. Journal} (2011)


\bibitem{EHLW10}
Escher J., Hillairet M., Lauren\c{c}ot Ph., Walker Ch.: 
\newblock Thin film equations with soluble surfactant and gravity:
modeling and stability of steady states, 
\newblock  Preprint (2010)


%\bibitem{Fal2010}
%Fallest D.W., Lichtenberger A.M., Fox C.J., Daniels K.E.:
%\newblock Fluorescent visualization of a spreading surfactant,
%\newblock {New J. Phys.}, 12, 073029 (2010)



\bibitem{Heinz}
Fehrenbach H.:
\newblock{Alveolar epithelial type II cell: defender of the alveolus revisited},
\newblock{Respiratory Research, 2(1),33 (2001)}

\bibitem{GW} Garcke H., Wieland S.: 
\newblock Surfactant spreading on thin viscous films: nonnegative solutions
of a coupled degenerate system, 
\newblock {SIAM J. Math. Anal.}, 37(6):2025?2048 (2006)


\bibitem{GG}
Gaver D.P., Grotberg J.B.: 
\newblock The dynamics of a localized surfactant on a thin film, 
\newblock { J. Fluid Mech.}, 213:127--148 (1990)



%\bibitem{location}
%Hage E.:
%\newblock{Electron microscopic identification of several types of endocrine cells in the bronchial epithelium of human foetuses}, 
%\newblock{Zeitschrift f\"ur Zellforschung und Mikroskopische Anatomie, 141(3):401 - 412 (1973)}

\bibitem{Haies}
Haies D.M., Gil J., and Weibel E.R.:
\newblock{Morphometric Study of Rat Lung Cells I. Numerical and Dimensional Characteristics of Parenchymal Cell Population}, 
\newblock{American Review of Respiratory Disease, 123(5):533-541 (1981)}

\bibitem{Halp1997}
Halpern D., Jensen O.E., Grotberg J.B.: 
\newblock A theoretical study of surfactant and liquid delivery into the lung, 
\newblock {Journal of Applied Physiology}, 85:333--352 (1998)

\bibitem{Halp2008}
Halpern D., Fujioka H., Takayama S., Grotberg J.B.: 
\newblock Liquid and surfactant delivery into pulmonary airways, 
\newblock {Respiratory Physiology and Neurobiology}, 163:222--231 (2008)


%\bibitem{GT}
%D. Gilbarg and N.S. Trudinger.
%\newblock {Elliptic Partial Differential Equations of Second Order.}
%Springer-Verlag, Berlin, 1977.

\bibitem{JG}
Jensen O.E., Grotberg J.B.: 
\newblock Insoluble surfactant spreading on a thin viscous
film: shock evolution and film rupture, 
\newblock {J. Fluid Mech.}, 240:259--288 (1992)


%\bibitem{JeGr93}
%Jensen O.E., Grotberg J.B.: 
%\newblock The spreading of heat or soluble surfactant along a thin liquid film, 
%\newblock {Phys. Fluids A}, 5:58--68 (1993)


%\bibitem{Jo1993}
%Jobe A.H.:  
%\newblock Pulmanory surfactant therapy, 
%\newblock { N. Engl. J. Med.}, 328:861--868 (1993)



\bibitem{kang2016} Kang D., Nadim A., and Chugunova M.:   Dynamics and equilibria of thin viscous coating films on a rotating sphere, Journal of Fluid Mechanics, 791, 495-518 (2016)

\bibitem{kang2017} Kang D., Nadim A., and Chugunova M.:   Marangoni effects on a thin liquid film coating a sphere with axial or radial thermal gradients, Physics of Fluids, 29(7), 072106 (2017)

\bibitem{sphere2}
Kang D.,  Sangsawang T.  and  Zhang J.:
\newblock{Weak  solution  of  a  doubly  degenerate  parabolic  equation,}
\newblock{arXiv preprint arXiv:1610.06303 (2016)}


%\bibitem{V1}
%Kamin S., Vazquez J.L.: 
%\newblock Asymptotic behaviour of solutions of the porous medium
%equation with changing sign, 
%\newblock {SIAM J. Math. Anal.}, 22(1):34--45 (1991)

%\bibitem{Lady} Lady{\v{z}}enskaja O.A., Solonnikov V.A.,   Ural'ceva N.N.: 
%\newblock {Linear and quasilinear equations of parabolic type}, 
%  Vol. 23, American Mathematical Society,
%Providence (1967)

%\bibitem{Levi2007}
%Levy R., Shearer  M., Witelski T.P.: 
%\newblock Gravity-driven thin liquid films
%with insoluble surfactant: smooth traveling waves, 
%\newblock{ European Journal of Applied Mathematics}, 18:679--708 (2007)
%
%\bibitem{Levi2006}
%Levy R., Shearer M.: 
%\newblock The motion of a thin liquid film driven by surfactant and gravity, 
%\newblock {SIAM Journal of Applied Mathematics}, 66:1588--1609 (2006)


%\bibitem{Lions} Lions J.L.: 
%\newblock {Quelques m\'{e}thodes de r\'{e}solution des probl\`{e}mes aux
%limites non lin\'{e}aires,} Dunod (1969)


\bibitem{Luc}
Lucassen J., Hansen R.S.: 
\newblock {Damping of Waves on Monolayer-Covered Surfaces
II. Influence of Bulk-to-Surface Diffusional Interchange on Ripple Characteristics}, 
\newblock {Journal of Colloid and Interface Science}, 23: 319--328 (1967)





\bibitem{Mason2006}
Mason R.J.: 
\newblock{Biology of alveolar type II cells}, 
\newblock{Respirology, 11:s1 (2006)}

%\bibitem{Matar02}
%Matar O.K.: 
%\newblock Nonlinear evolution of thin free viscous films
%in the presence of soluble surfactant, 
%\newblock  { Phys. Fluids}, 14:4216--4234 (2002)

%\bibitem{Matar2009}
%Matar O.K., Craster R.V.: 
%\newblock Dynamics of surfactant-assisted spreading, 
%\newblock { Soft Matter}, 5(20):3801--3809 (2009)

%\bibitem{MSL93} W.J.~Milliken, H.A.~Stone, L.G.~Leal.
%\newblock The effect of surfactant on the transient motion of Newtonian drops.
%\newblock  {Phys. Fluids~A}, 5:69--79, 1993.

%\bibitem{N1}
%L.~Nirenberg.
%\newblock An extended interpolation inequality.
%\newblock {\em Ann. Scuola Norm. Sup. Pisa (3)}, 20:733--737, 1966.


\bibitem{Michael}
Matthay M.A., Folkesson H.G., Clerici C.: 
\newblock{Lung Epithelial Fluid Transport and the Resolution of Pulmonary Edema}, 
\newblock{Physiological Reviews, Vol. 82 no. 3, 569-600  (2002)}

\bibitem{nadim}
Nadim A.: 
\newblock A Concise Introduction to Surface Rheology with Application to Dilute Emulsions of Viscous Drops, 
\newblock { Chem.\ Engr.\ Commun.}, 148--150: 391--407 (1996)



%\bibitem{Not2000}
%Notter  R.H.: 
% \newblock Lung Surfactants: Basic Science and Clinical Applications, 
% \newblock  Marcel Dekker, New York (2000)

\bibitem{Frans}
Notter R.H., Gupta R., Schwan A.L., Wang Z., Shkoor M.G.,  Walther F.J., 
\newblock{Synthetic lung surfactants containing SP-B and SP-C peptides plus novel phospholipase-resistant lipids or glycerophospholipids},
\newblock{ PeerJ, 4, e2635 (2016)}

\bibitem{nina}
Nouraeyan N., Lambrinakos-Raymond A., Leone M., and Sant'Anna G.: 
\newblock{Surfactant administration in neonates: A review of delivery methods}, 
\newblock{Canadian Journal of Respiratory Therapy, 50(3): 91--95 (2014)}

\bibitem{Elisa}
Parra E., Kinoshita K., and Needham D.: 
\newblock{Micropipette Technique Study of Natural and Synthetic Lung Surfactants at the Air-Water Interface:
Presence of a SP-B Analog Peptide Promotes Membrane Aggregation, Formation of Tightly Stacked Lamellae, and Growth of Myelin Figures}, 
\newblock{Langmuir, 32(41):10570 - 10581 (2016)}


%\bibitem{Pet2011}
%Peterson E.R., Shearer M.: 
%\newblock Radial Spreading of a Surfactant on a Thin Liquid Film, 
%\newblock {Appl Math Res Express}, 2011(1):1--22 (2011)


%\bibitem{Renardy96a}
%Renardy M.: 
%\newblock  A singularly perturbed problem related to surfactant
%spreading on thin films, 
%\newblock { Nonlinear Anal.}, 27:287--296 (1996)
%
%\bibitem{Renardy96b}
%Renardy M.: 
%\newblock On an equation describing the spreading of surfactants on thin films,
%\newblock { Nonlinear Anal.}, 26:1207--1219 (1996)

%\bibitem{Renardy97}
%Renardy M.: 
%\newblock A degenerate parabolic-hyperbolic system modeling the spreading of
%surfactants, 
%\newblock {SIAM J. Math. Anal.}, 28:1048--1063 (1997)

%\textcolor{red} { \bibitem{Ros}
%Milton J. Rosen.
%\newblock{Surfactants and interfacial phenomena}, 3rd ed.;
%John Wiley and Sons: New York, 2004.}


\bibitem{Shier2007}
Shier D., Butler J. and Lewis R.: 
\newblock{Human Anatomy and Physiology}
\newblock{ 11th Edition, McGraw-Hill (2007)}

\bibitem{Sim} Simon J.: 
\newblock Compact sets in the space $L^p(0,T; B)$, 
\newblock{Ann. Mat. Pura Appl.}, 146 (4):65-96 (1987)

%\bibitem{Sol} Solonnikov V.A.: 
%\newblock On boundary value problems for linear parabolic systems of
%differential equations of general form, 
%\newblock{Trudy Mat. Inst. Steklov}, 83:3--163 (1965)


\bibitem{sphere1}
Taranets R.M.: 
\newblock{Strong solutions of the thin film equation in spherical geometry}, 
\newblock{arXiv preprint arXiv:1709.10496} (2017)

%\bibitem{Touh}
%Touhami Y., Rana D., Neale G.H., Hornof V.: 
%\newblock  Study of polymer-surfactant interactions via surface tension measurements, 
%\newblock {Colloid and Polymer Science}, 279(3): 297--300 (2001)


%\bibitem{Tray1977}
%Traykov T.T., Ivanov I.B.: 
%\newblock Hydrodynamics of thin liquid films. Effects of surfactants
%on the velocity of thinning of emulsion films, 
%\newblock {Int. Z Multiphase Flow}, 3:471--483 (1977)

%\bibitem{V}
%O.V. Voinov.
%\newblock Hydrodynamics of wetting.
%\newblock {\em Fluid Dynamics}, 11(5):714--721, 1976.

%\bibitem{Vaz} V\'{a}zquez J.L.: 
%\newblock The porous medium equation: Mathematical theory, 
%\newblock  Oxford University Press (2007)

\bibitem{Verkman}
Verkman A.S., Matthay M.A., and Song Y.: 
\newblock{Aquaporin water channels and lung physiology}, 
\newblock{American Journal of Phisiology: Lung Cellura and Molecular Phisiology, 278(5):L867-L879 (2000)}

\bibitem{walther2014}
Walther F.J., Hernï¿œndez-Juviel J.M., Waring A.J.:  
\newblock{Aerosol delivery of synthetic lung surfactant,} 
PeerJ 2: e403  (2014)

\bibitem{space1}
West, J.B., Elliott, A.R., Guy, H.J., and Prisk, G.K.:    Pulmonary function in space, Jama, 277(24), 1957-1961 (1997)

\bibitem{Williams}
Williams M.C., Cao Y., Hinds A., Rishi A.K., and Wetterwald A.: 
\newblock{T1 alpha protein is developmentally regulated and expressed by alveolar type I cells, choroid plexus, and ciliary epithelia of adult rats,} 
\newblock{American Journal of Respiratory Cell and Molecular Biology, 14(6):577 - 585 (1996)}

%\bibitem{Wit2006}
%Witelski T.P., Shearer M., Levy R.: 
%\newblock  Growing surfactant waves in thin liquid films driven by gravity, 
%\newblock  {Applied Mathematics Research Express}, 15487:1--21 (2006)

%\bibitem{Zei}
%\textcolor{red}{E. Zeidler. Nonlinear functional analysis and its applications IIB.
%Springer-Verlag, New York, 1990.}

\end{thebibliography}


\end{document}